\documentclass[%
 preprint,
superscriptaddress,
 amsmath,amssymb,
 aps,
pra,
]{revtex4-1}

\usepackage{graphicx}
\usepackage{dcolumn}
\usepackage{bm}

\usepackage{color}

\usepackage{xcolor}

\usepackage{graphicx}
\usepackage{tikz}
\usetikzlibrary{arrows}
\newcommand{\ket}[1]{$\left|#1\right\rangle$}

\begin{document}

\title{Quantum theory of superfluorescence based on two-point correlation functions}

\author{Andrei Benediktovitch}
\email{andrei.benediktovitch@desy.de}
\affiliation{Center for Free-Electron Laser Science, DESY, Hamburg 22607, Germany}

\author{Vinay P. Majety}
\altaffiliation[Present address: ]{Indian Institute of Technology Tirupati, Tirupati, India}
\affiliation{Max Planck Institute for the Structure and Dynamics of Matter, Hamburg 22607, Germany}

\author{Nina Rohringer}
\affiliation{Center for Free-Electron Laser Science, DESY, Hamburg 22607, Germany}
\affiliation{Department of Physics, Universit{\"a}t Hamburg, Hamburg 22761, Germany}


\begin{abstract}
Irradiation of a medium by short intense pulses from x-ray / XUV free electron lasers can result in saturated photoionization of inner electronic shells. As a result an inversion of populations between core levels appears. The resulting fluorescent radiation can be amplified during its propagation through the inverted medium and results in intense, quasi transform-limited radiation bursts. While the optical counterpart of this phenomena, known as superfluorescence, was intensively investigated, a generalized treatment is needed in the x-ray / XUV domain, where the dynamics of pumping and evolution due to fast decay processes play a crucial role. To provide a general theoretical approach, we start from the fundamental, quantized minimal coupling Hamiltonian of light-matter interaction and after a series of approximations arrive at a closed system of equations for the two-point correlation function of atomic coherences and the two-time correlation function of the emitted field. The obtained formalism enables us to investigate collective spontaneous emission in various regimes. It is extended consistently to include incoherent processes that are relevant in the x-ray / XUV domain. These processes are introduced into the formalism by corresponding Lindblad superoperators. The connection to other approaches is discussed and numerical examples related to recent experiments are presented.


\end{abstract}


\maketitle


\section{\label{sec: Introduction} Introduction}

The interaction of x-rays with matter has been providing unique information about the structure of matter at Angstrom length-scale for more than a century. One of the reasons for its success is the weak interaction between x-rays and atomic systems: the overwhelming majority of x-ray laboratory and synchrotron experiments can be described within linear response approaches. Recently, however, the advent of x-ray free electron lasers (XFEL) has opened up exciting new possibilities for the investigation of x-ray - matter interaction. Strongly nonlinear x-ray driven processes have become experimentally reachable. In particular, this can manifest itself in the stimulated emission of radiation that was demonstrated in gases \cite{2012'Rohringer}, \cite{2013'Weninger}, \cite{2018'Mercadier}, solids \cite{2015'Yoneda} and solutions \cite{2018'Kroll}. The emitted bursts of radiation were shown to have large photon numbers and a bandwidth much narrower than the pumping XFEL pulse. Hence they are promising both for stimulated x-ray spectroscopy \cite{2018'Kroll}, \cite{2013'Weninger} and for use as dedicated x-ray photon sources with unique characteristics that can be employed for further (pump-probe) experiments.

The process described above has been referred to as x-ray lasing in many cases \cite{2012'Rohringer}, \cite{2015'Yoneda}, \citep{2014'Weninger}. The population inversion is obtained as result of inner-shell photo-ionization or Auger decay following the inner-shell ionization. The shape of the gain region is determined by the overlap of the pumping x-ray beam and the driven medium. Focusing the x-rays into a target results in an elongated, cylindrical shape. Comparing the process to traditional lasing in the optical regime, there are a number of marked differences: i) the x-ray pumping is applied in form of a short pulse propagating through the medium, ii) there is no feedback system employed, iii) there can be numerous loss and decay processes that act on timescales comparable to the emission process, resulting in transient gain. Each emission starts from noise-like spontaneous radiation. The portion of radiation emitted along the axis of the beam is amplified during a single-pass propagation through the inverted medium and, if the number of involved inverted atoms is large enough, the stored energy is radiated within a short, intense pulse. A comparable phenomenon in the optical domain is referred to as superfluorescence, and has seen extensive experimental investigation, see Ref. \cite{1973'Skribanowitz}, \cite{1977'Gibbs}, \cite{1979'Vrehen}, \cite{1987'Malcuit} for early examples. Notably, the term superfluorescence is often used synonymously with superradiance. Furthermore, the crossover from superfluorescence to amplified spontaneous emission is not sharp \cite{1987'Malcuit}. Hence, theory becomes essential for the interpretation of experimental results.

Theoretical modeling of the phenomena described above, which we will generally refer to as superfluorescence, is an involved problem and has been  studied extensively (cf. Ref. \cite{1954'Dicke}, \cite{1982'Gross}, \cite{book'superradiance}, \cite{2017'Kocharovsky} and references therein). The complexity of the problem was described in Ref. \cite{1982'Gross}, p. 336, as follows: ``the problem of superradiance in free space is in general a problem of three-dimensional non-linear diffraction theory, further complicated by the quantum nature of the radiated field at the early stage of the emission process.'' Hence theoretical models used to describe experimental data typically have to rely on a number of approximations in order to reduce the complexity of the problem. The range of validity is conditioned by the parameters of the system under study. In many cases, a one-dimensional semi-classical Maxwell-Bloch model is used for the description of x-ray / XUV experiments based on XFEL pumping \cite{2013'Weninger}, \cite{2018'Kroll}, \cite{2013'Kimberg}, \cite{2014'Weninger}, \cite{2018'Lyu} or plasma-based pumping \cite{2000'Larroche}, \cite{2015'Depresseux}, \cite{2012'Oliva}, \cite{2014'Wang}, \cite{2015'Oliva}. The reduction to a one-dimensional problem is justified by the shape of the excited medium, which resembles an extensively elongated cylinder. The non-linear nature of the stimulated process can also be captured in the Maxwell-Bloch formalism. However, while having a clear interpretation and being comparatively easy to simulate numerically, the semi-classical 1D Maxwell-Bloch equations are not able to describe the initial stage of the amplification process. Namely, the semi-classical 1D Maxwell-Bloch equations form a \emph{homogeneous} system of differential equations. Thus if all atoms are initially in the inverted state and neither field nor polarization is present at that moment, there will also not be any field or polarization at later times. Here, accounting for the quantum nature of the radiated field is an unavoidable ingredient as only this will allow the system to relax by means of spontaneous emission. 

Several approaches have been proposed to extend the semi-classical Maxwell-Bloch formalism in order to account for spontaneous emission. One way is to assume a fixed, small value of the polarization at the initial time. In terms of a Bloch sphere representation, this corresponds to the introduction of a small tipping angle af the Bloch vector from North pole. In the 1970-s, multiple  ways to calculate the actual value of this tipping angle were proposed \cite{1971'Rehler}, \cite{1975'BonifacioI}, \cite{1975'BonifacioII}, \cite{1976'MacGillivray}, \cite{1979'Polder}; in Ref.~\cite{1979'Vrehen}, these approaches were compared and an experimental confirmation for the values obtained in Ref.~\cite{1975'BonifacioI}, \cite{1979'Polder} was given. However, even though such a tipping angle approach can be sufficient to reproduce the general shape of the emitted pulses, some important features of the superfluorescent emission are missing in this approach. In particular, it fails to account for the strong fluctuations in shape and arrival times that superfluorescent pulses exhibit as a consequence of their origin in quantum fluctuations. These can be described more adequately in a framework such as proposed in Ref.~\cite{1979'Haake}. There, the observables are obtained from an ensemble of solutions to the Maxwell-Bloch equations corresponding to random initial polarizations. The distribution of the initial polarization is obtained from properties of the polarization operator assuming all atoms to be in the excited state. In this approach, the process that excites the atoms is explicitly separated from the superfluorescence dynamics. However, for x-ray / XUV experiments the timescales for processes initiated by pumping and the timescale for the superfluorescence dynamics can in general be of the same order. The simultaneous description of evolution due to pumping and superfluorescent emission can be achieved, if random noise terms are considered in the equations for polarization or polarization and field instead of random initial conditions. These noise terms have been  derived from first principles using phase space approaches in Ref.~\cite{1989'Maki}, \cite{1991'Drummond} and give contributions to both the polarization and the emitted field. The noise has a multiplicative structure and --- in general --- can lead to unstable numerical behavior of simulations after finite propagation times. In paper \cite{2000'Larroche}, the noise terms for the polarization were derived semi-phenomenologically. Here, we avoid the phenomenological introduction of terms into the Maxwell-Bloch equations and instead start from a quantum-mechanical Hamiltonian. Following a series of approximations, we arrive at equations of motion for the correlation function of atomic coherences and the field correlation function. In this formulation, the spontaneous and stimulated emission enter on the same footing, hence the crossover between them can be analyzed straightforwardly.

The paper is organized as follows: in section \ref{sec: Derivation_I}, we start from the basic Hamiltonian that describes the interaction of two-level atoms with the electromagnetic field arriving at a system of integro-differential equations for the level populations, the correlation function of atomic coherences and the correlation function of the field. In section \ref{sec: Modification}, these equations are generalized to include a number of incoherent processes, which are introduced by means of Lindblad superoperators. In section \ref{sec: Numerical examples}, we discuss the crossover to spontaneous emission and to the semi-classical Maxwell-Bloch  equations. Futhermore, we compare our approach to the Maxwell-Bloch equations with stochastic noise terms and present a modeling example of a realistic system. Details of the calculation that were skipped in the main text are presented in Appendices \ref{appendix: derivation H-L} -- \ref{appendix: Estimation of noise-term correlation factor for Maxwell-Bloch equations}. SI units are used throughout this article.

\section{\label{sec: Derivation_I} Derivation of evolution equations of the correlation function }

In order to study the fundamental features of collective spontaneous emission, we employ the following model. Consider an elongated, roughly  cylindrical medium consisting of a large number of two-level atoms. We assume an ideal swept-gain pumping of the system, that is a population inversion is created quasi instantaneously by a pump-pulse front that transverses the medium with speed of light \cite{1975'Hopf}, \cite{1975'Hopf_q}, \cite{1975'Louisell}. The temporal evolution of the atomic ensemble in interaction with (vacuum) electromagnetic field modes is determined by the well-known Hamiltonian \citep{1997'Scully}:

\begin{align}
\label{eqs: H}
\hat{H}&=\sum_{a}\hbar\Omega\hat{\sigma}_{z}^{(a)}+\int d^3 \vec{k}\sum_{s}\hbar\omega_{\vec{k}}\hat{a}_{\vec{k},s}^{\dagger}\hat{a}_{\vec{k},s}+\int d^3 \vec{k}\sum_{a,s}\hbar(g_{\vec{k},s}e^{i\vec{k}\vec{r}_{a}}\hat{\sigma}_{+}^{(a)}\hat{a}_{\vec{k},s}+g_{\vec{k},s}^*e^{-i\vec{k}\vec{r}_{a}}\hat{a}_{\vec{k},s}^{\dagger}\hat{\sigma}_{-}^{(a)}).
\end{align}

\noindent Here, $\Omega$ denotes the frequency of the transition between the ground \ket{g} and excited \ket{e} state of each atom. The operators $\hat{\sigma}_{z,+,-}^{(a)}$ correspond to observables of two-level atom $a$: $\hat{\sigma}_{z}$ corresponds to the population inversion

\begin{align}
\hat{\sigma}_{z} = \frac{1}{2}(|e \rangle \langle e | - |g \rangle \langle g |)
\end{align}

\noindent and $\hat{\sigma}_{+,-}$ correspond to atomic coherences

\begin{align}
\label{eqs: def Spm}
\hat{\sigma}_{+} = |e \rangle \langle g |, \hat{\sigma}_{-} = |g \rangle \langle e |.
\end{align}

\noindent The operators $\hat{a}_{\vec{k},s}^{\dagger},\hat{a}_{\vec{k},s}$ are creation and annihilation operators for photons in the electromagnetic field mode with wavevector $\vec k$, frequency $\omega_{\vec{k}}$ and polarization $s$; $\vec{r}_{a}$ gives the position of atom $a$ and $g_{\vec{k},s}$ is the coupling constant for an atom and the electromagnetic field due to the $\vec A \cdot \vec p$ interaction term. It is given as

\begin{align}
\label{eqs: gks and p}
g_{\vec{k},s}&=\frac{e}{m}\sqrt{\frac{1}{2(2\pi)^{3}\epsilon_{0}\hbar\omega_{\vec{k}}}}\vec{p} \cdot \vec{e}_{\vec{k},s},\;\;\; \text{with} \;\;\; \vec{p}= \langle e |\hat{\vec{p}}| g \rangle,
\end{align}

\noindent where $\vec{e}_{\vec{k},s}$ is a unit polarization vector. Notably, the interaction Hamiltonian in Eq.~(\ref{eqs: H}) implies the rotating wave approximation. Furthermore, the dipole approximation was invoked in Eq.~(\ref{eqs: gks and p}),which is valid for core-shell excitations in the x-ray domain, see Ref.~\cite{2006'Altarelli}.

\subsection{Heisenberg - Langevin equations}

Our aim is to obtain a closed system of equations for the atomic and field observables. In order to accomplish this, we need the generic equation of motion for these quantities first. Following the usual derivation of Heisenberg - Langevin equations \cite{1997'Scully}, \cite{1974'Agarwal}, we arrive at (see the detailed derivation in Appendix \ref{appendix: derivation H-L}):

\begin{align}
\label{eqs: H-L general Sm}
\frac{d\hat{\sigma}_{-}^{(a)}(t)}{dt} & =-i\Omega\hat{\sigma}_{-}^{(a)}(t)-\frac{\Gamma_{sp}}{2}\hat{\sigma}_{-}^{(a)}(t)+\frac{2 i e}{m \hbar c}\hat{\sigma}_{z}^{(a)}(t)\vec{p}\cdot\vec{\hat{A}}_{+}^{(a)}(t) + \hat{F}_{-}^{(a)}(t),\\
\label{eqs: H-L general Sz}
\frac{d\hat{\sigma}_{z}^{(a)}(t)}{dt} & =-\Gamma_{sp}\hat{\sigma}_{+}^{(a)}(t)\hat{\sigma}_{-}^{(a)}(t)+\frac{i e}{m \hbar c} \left[ \vec{p}^*\cdot\vec{\hat{A}}_{-}^{(a)}(t)\hat{\sigma}_{-}^{(a)}(t)-\hat{\sigma}_{+}^{(a)}(t)\vec{p}\cdot\vec{\hat{A}}_{+}^{(a)}(t)\right] +\hat{F}_{z}^{(a)}(t),\\ 
\label{eqs: H-L general A}  
\vec{\hat{A}}_{+}^{(a)}(t)&=-\frac{iec}{16\pi^{3}m\epsilon_{0}}\int d^3 \vec{k}\sum_{s}\frac{1}{\omega_{\vec{k}}}\vec{e}_{\vec{k},s}(\vec{e}_{\vec{k},s}^{*}\cdot\vec{p}^{*})\int_{0}^{t}dt'\sum_{b\neq a}e^{i(\vec{k}(\vec{r_{a}}-\vec{r_{b}})-\omega_{\vec{k}}(t-t'))}\hat{\sigma}_{-}^{(b)}(t').
\end{align}

\noindent Here, $\vec{\hat{A}}_{+}^{(a)}$ is the positive-frequency part of the vector potential at the position of atom $a$ due to all other atoms, $\hat{F}_{z,\pm}^{(a)}(t)$ are stochastic Langevin terms due to interaction with vacuum field (see the details in Appendix \ref{appendix: derivation H-L}), $\Gamma_{sp}=\frac{e^2\Omega |\vec p|^2}{3\pi\epsilon_0\hbar m^2 c^3}$ is the spontaneous emission rate.

Up to this point, no approximations were made regarding the emitted electromagnetic field. Due to this generality, Eqs. (\ref{eqs: H-L general Sm}) - (\ref{eqs: H-L general A}) are non-Markovian with respect to interaction among the atoms: the evolution of a given atom $a$ is conditioned by operator values of all other atoms at all previous points in time. These operator equations, however, are still of limited use for treating a large number of atoms. In order to proceed, let us make the following approximations: i) omit the polarization properties of the radiation and ii) restrict the description of the field to one dimension. The first approximation is equivalent to the assumption that all atoms have the vector of their dipole transition matrix element oriented along the same (fixed) direction. This direction is assumed to be orthogonal to the propagation direction of the pump field and is further considered as the polarization direction of the induced field. This assumption is not crucial for the developed formalism, but it essentially simplifies the equations. In order to perform the reduction to a 1D model let us take into account that only those modes of the field that have their wavevectors roughly aligned with the axis of the cylindrical medium will interact with large number of atoms and thus will be amplified. Based on this consideration, let us assume that all waves with a wavevector oriented along the system $z$ axis within a small solid angle $\Delta o$ have the same magnitude. Waves with a wavevector outside $\Delta o$ are neglected. In particular, waves are neglected that would be counterpropagating to the swept pumping field. 
These assumptions correspond to the following way of calculating the integral over wavevectors in (\ref{eqs: H-L general A}):

\begin{align}
\label{eqs: 1D approximation}
\int d^3 \vec{k}&\rightarrow\Delta o\int_{0}^{\infty}d\omega\frac{\omega^{2}}{c^{3}}.
\end{align}

The actual value of $\Delta o$ to be used in calculations can be estimated from geometrical considerations as $\Delta o_g \sim \pi R^2/L^2$ where $R$ is radius of the medium and $L$ is its length. However, if $\Delta o_g$ exceeds the solid angle $\Delta o_d \sim \lambda^2/R^2$, diffraction effects become important and one can expect that the emission decomposes into independently radiating regions of size $\Delta o_d$, see Ref.~\cite{1982'Gross} for further discussion.

With help of (\ref{eqs: 1D approximation}), performing similar steps as in the derivation of (\ref{eqs: Sm sp}), and omitting the vectorial properties of the field one obtains

\begin{align}
\label{eqs: A in t}
\hat{A}_{+}^{(a)}(t)& =-\frac{i\Delta o e p^* \Omega}{8\pi^{2}\epsilon_{0} m c^{2}}\int_{0}^{t}dt'\sum_{b \neq a}\delta\left[t-\frac{z_{a}}{c}-(t'-\frac{z_{b}}{c})\right]\hat{\sigma}_{-}^{(b)}(t').
\end{align}

\noindent Due to the assumption of swept-gain pumping, which  propagates with speed $c$ through the medium, the non-trivial evolution of atom $a$  starts at the time $z_a/c$. Hence, all atomic operators $\hat{\sigma}^{(a)}$ have a time dependence of the form $\hat{\sigma}^{(a)}(t-\frac{z_{a}}{c})$. Let us change variables to the retarded time $\tau$, defined for each atom $a$ as $\tau=t-\frac{z_{a}}{c}$. In terms of this variable one obtains from Eq.~(\ref{eqs: A in t}) 

\begin{align}
\label{eqs: A from S}
\hat{A}_{+}^{(a)}(\tau) & =-\frac{i \Delta o e p \Omega}{16\pi^{2}\epsilon_{0} m c^{2}}\sum_{b<a}\hat{\sigma}_{-}^{(b)}(\tau).
\end{align}

\noindent Here, $\sum_{b<a}$ means that one has to sum over atoms $b$ positioned before (with respect to pump propagation) the considered atom $a$. With the help of (\ref{eqs: A from S}) the expressions (\ref{eqs: H-L general Sm})-(\ref{eqs: H-L general Sz}) take the form

\begin{align}
\label{eqs: HL Sm}
\frac{d\hat{\sigma}_{-}^{(a)}(\tau)}{d\tau}&=-i\Omega\hat{\sigma}_{-}^{(a)}(\tau)-\frac{\Gamma_{sp}}{2}\hat{\sigma}_{-}^{(a)}(\tau)+\frac{3\text{\ensuremath{\Delta o}}}{8\pi}\Gamma_{sp}\sum_{b<a}\hat{\sigma}_{z}^{(a)}(\tau)\hat{\sigma}_{-}^{(b)}(\tau)+\hat{F}_{-}^{(a)}(\tau), \\
\label{eqs: HL Sz}
\frac{d\hat{\sigma}_{z}^{(a)}(\tau)}{d\tau}&=-\Gamma_{sp}\hat{\sigma}_{+}^{(a)}(\tau)\hat{\sigma}_{-}^{(a)}(\tau)-\frac{3\text{\ensuremath{\Delta o}}}{16\pi}\Gamma_{sp}\sum_{b<a}\left[\hat{\sigma}_{+}^{(a)}(\tau)\hat{\sigma}_{-}^{(b)}(\tau)+\hat{\sigma}_{+}^{(b)}(\tau)\hat{\sigma}_{-}^{(a)}(\tau)\right]+\hat{F}_{z}^{(a)}(\tau).
\end{align}

\subsection{Equations for atomic observables}

The operator equations (\ref{eqs: HL Sm})-(\ref{eqs: HL Sz}) can be used to obtain equations for their  mean values. In the following we will obtain equations for the expectation values of the atomic coherences, population inversion and we will also introduce a correlation function of atomic coherences.

\subsubsection*{Coherences}

The mean value of atomic coherences is given by $\langle\hat{\sigma}_{-}^{(a)}\rangle=\text{Tr}(\hat{\sigma}_{-}^{(a)}\hat{\rho}^{(a)})=\rho_{eg}^{(a)},\langle\hat{\sigma}_{+}^{(a)}\rangle=\rho_{eg}^{(a)*}$, where $\hat{\rho}^{(a)}$ denotes the one-atom density matrix for atom $a$. Assuming factorization of the operators acting on different atoms (this is valid at the initial moment of time) and taking the mean value of equation (\ref{eqs: HL Sm}), we end up with a homogeneous equation for each $\langle\hat{\sigma}_{-}^{(a)}\rangle$. If initially there is no coherent excitation in the system $\langle\hat{\sigma}_{-}^{(a)}(\tau=0)\rangle=0$, it will remain the same during the evolution:

\begin{align}
\label{eqs: mean Sm}
\langle\hat{\sigma}_{\pm}^{(a)}(\tau)\rangle & =0.
\end{align}

\subsubsection*{Population inversion}

We define the population inversion to be $\langle\hat{\sigma}_{z}^{(a)}\rangle=\frac{1}{2}(\rho_{ee}^{(a)}-\rho_{gg}^{(a)})$, where $\rho_{ee}^{(a)}$ and $\rho_{gg}^{(a)}$ are the populations of the ground and excited states of atom $a$. 
From (\ref{eqs: HL Sz}) one obtains 

\begin{align}
\label{eqs: mean Sz}
\frac{d\langle \hat{\sigma}_{z}^{(a)}(\tau)\rangle}{d\tau}=-\Gamma_{sp}\rho_{ee}^{(a)}(\tau)-\frac{3\text{\ensuremath{\Delta o}}}{16\pi}\Gamma_{sp}\sum_{b<a}\left[\langle\hat{\sigma}_{+}^{(a)}(\tau)\hat{\sigma}_{-}^{(b)}(\tau)\rangle+\langle\hat{\sigma}_{+}^{(b)}(\tau)\hat{\sigma}_{-}^{(a)}(\tau)\rangle\right].
\end{align}

\noindent Here, we have taken into account the property $\langle\hat{\sigma}_{+}^{(a)}(\tau)\hat{\sigma}_{-}^{(a)}(\tau)\rangle  =\rho_{ee}^{(a)}(\tau)$ of the coherence operators, which follows immediately from (\ref{eqs: def Spm}). The noise term $\hat{F}_{z}^{(a)}(\tau)$ has disappeared due to the property (\ref{eqs: Fz}). Hence, the equation (\ref{eqs: mean Sz}) is a non-stochastic $c$-number differential equation, as well as other equations for observables.

Notably, equation (\ref{eqs: mean Sz}) involves the quantity $\langle\hat{\sigma}_{+}^{(a)}(\tau)\hat{\sigma}_{-}^{(b)}(\tau)\rangle$, which we shall subsequently  consider in more detail.

\subsubsection*{Correlation function of atomic coherences}

The value of $\langle\hat{\sigma}_{+}^{(a)}(\tau)\hat{\sigma}_{-}^{(b)}(\tau)\rangle$ quantifies the joint probability for atoms $a$ and $b$ to exhibit coherence. It is related to the two-atom density matrix $\hat{\rho}^{(ab)}$ as

\begin{align}
\label{eqs: meaning of SpSm}
\langle\hat{\sigma}_{+}^{(a)}(\tau)\hat{\sigma}_{-}^{(b)}(\tau)\rangle = \text{Tr}\left(\hat{\rho}^{(ab)}(\tau)\hat{\sigma}_{+}^{(a)}(\tau)\hat{\sigma}_{-}^{(b)}(\tau)\rangle\right) = \rho_{g_{a}e_{b},e_{a}g_{b}}^{(ab)}.
\end{align}

\noindent Similarly to the quantity $\langle\hat{E}_{-}(\vec{r}_1, t_1)\hat{E}_{+}(\vec{r}_2, t_2)\rangle = \text{Tr}\left(\hat{\rho}^{(F)}\hat{E}_{-}(\vec{r}_1, t_1)\hat{E}_{+}(\vec{r}_2, t_2)\right)$ ---being named correlation function of the field \cite{1997'Scully}, \cite{1963'Glauber} we can refer to (\ref{eqs: meaning of SpSm}) as the correlation function of atomic coherences. Since the two-level atom polarization is proportional to the atomic coherence, the quantity (\ref{eqs: meaning of SpSm}) gives the correlation of atomic polarizations up to a factor $|p|^2$.

The equation of motion of (\ref{eqs: meaning of SpSm}) can be obtained based on (\ref{eqs: HL Sm}), see the details in Appendix \ref{appendix: Equation of motion for product of operators}:

\begin{align}
\label{eqs: mean SpSm v0}
\frac{d\langle\hat{\sigma}_{+}^{(a)}(\tau)\hat{\sigma}_{-}^{(b)}(\tau)\rangle}{d\tau}&=-\Gamma_{sp}\langle\hat{\sigma}_{+}^{(a)}(\tau)\hat{\sigma}_{-}^{(b)}(\tau)\rangle\\ \nonumber
+&\frac{3\text{\ensuremath{\Delta o}}}{8\pi}\Gamma_{sp}\left(\sum_{c<a}\langle\hat{\sigma}_{+}^{(c)}(\tau)\hat{\sigma}_{z}^{(a)}(\tau)\hat{\sigma}_{-}^{(b)}(\tau)\rangle+\sum_{c<b}\langle\hat{\sigma}_{+}^{(a)}(\tau)\hat{\sigma}_{z}^{(b)}(\tau)\hat{\sigma}_{-}^{(c)}(\tau)\rangle\right).
\end{align}

\noindent Equations (\ref{eqs: mean Sz}) and (\ref{eqs: mean SpSm v0}) do not yet form closed system of equations, because of the triple operator products on the r.h.s. of (\ref{eqs: mean SpSm v0}). It should be noted that similar equations for triple operator products would couple to quartic operator product and so on. The hierarchy of these equations would reflect the BBGKY hierarchy of reduced density matrix equations \cite{2016'Bonitz}. To obtain a finite and closed system of equations, we factorize triple operator products as

\begin{align}
\label{eqs: approx triple factr}
\langle\hat{\sigma}_{+}^{(a)}(\tau)\hat{\sigma}_{z}^{(b)}(\tau)\hat{\sigma}_{-}^{(c)}(\tau)\rangle\underset{b\neq a,c}{\approx}\langle\hat{\sigma}_{z}^{(b)}(\tau)\rangle\langle\hat{\sigma}_{+}^{(a)}(\tau)\hat{\sigma}_{-}^{(c)}(\tau)\rangle.
\end{align}

\noindent The triple product factorization generally enables the description of a large class of phenomena and form the basis for most important kinetic equations \cite{2016'Bonitz}. With respect to superfluorescence type of problems, it was demonstrated numerically that dropping third-order cumulants (equivalent to the factorization (\ref{eqs: approx triple factr})) does not affect the solution of the steady-state superradiance problem \cite{2010'Meiser_1}, \cite{2010'Meiser_2}. Performing the factorization (\ref{eqs: approx triple factr}) and treating  the terms with $c=b$ or $c=a$ on the r.h.s. of (\ref{eqs: mean SpSm v0}) separately, we obtain

\begin{align}
\label{eqs: mean SpSm}
\frac{d\langle\hat{\sigma}_{+}^{(a)}(\tau)\hat{\sigma}_{-}^{(b)}(\tau)\rangle}{d\tau}&=-\Gamma_{sp}\langle\hat{\sigma}_{+}^{(a)}(\tau)\hat{\sigma}_{-}^{(b)}(\tau)\rangle\\ \nonumber
+&\frac{3\Delta o}{8\pi}\Gamma_{sp}\left(\sum_{c<a,c\neq b}\langle\hat{\sigma}_{z}^{(a)}(\tau)\rangle\langle\hat{\sigma}_{+}^{(c)}(\tau)\hat{\sigma}_{-}^{(b)}(\tau)\rangle + \sum_{c<b,c\neq a}\langle\hat{\sigma}_{z}^{(b)}(\tau)\rangle\langle\hat{\sigma}_{+}^{(a)}(\tau)\hat{\sigma}_{-}^{(c)}(\tau)\rangle\right)\\ \nonumber
+&\frac{3\Delta o}{8\pi}\Gamma_{sp}\left(\langle\hat{\sigma}_{z}^{(a)}(\tau)\rangle\rho_{ee}^{(b)}(\tau)\Theta(z_a-z_b)+\langle\hat{\sigma}_{z}^{(b)}(\tau)\rangle\rho_{ee}^{(a)}(\tau)\Theta(z_b-z_a)\right),
\end{align}

\noindent where $\Theta(z)$ is the Heaviside function.
Equations (\ref{eqs: mean Sz}) and (\ref{eqs: mean SpSm}) form a closed set of equations for the evolution of the electronic degrees of freedom.

\subsection{Equations for field observables}

In most of the experiments dealing with x-ray / XUV collective emission, the measured observables are related to the properties of the emitted electromagnetic field. Hence, we consider in more detail properties of the emitted field.  

\subsubsection*{Slowly varying amplitude of the vector potential}

The vector potential is given by (\ref{eqs: A from S}); due to (\ref{eqs: mean Sm}) one straightforwardly obtains

\begin{align}
\langle\hat{A}_{\pm}^{(a)}\rangle & =0.
\end{align}

This implies that the ensemble average emitted field cannot be described by a classical amplitude. Instead, the description should be given in terms of statistical properties of the field.

\subsubsection*{Intensity}

The statistical property of primary interest is the intensity of the emitted field. Let us quantify this by the number of photons emitted 
per solid angle $\Delta o$ per cross section of the system per unit time. Thus, for the intensity at the position of atom $a$, we can take the value of the Poynting vector and normalize by $\Delta o\hbar\omega$. Applying normal ordering, we obtain

\begin{align}
\label{eqs: I za}
I^{(a)}(\tau) & =\frac{2\epsilon_{0}\omega}{\Delta o\hbar c}\langle\hat{A}_{-}^{(a)}(\tau)\hat{A}_{+}^{(a)}(\tau)\rangle=\frac{3\Delta o}{32\pi\lambda^{2}}\Gamma_{sp}\sum_{b<a,c<a}\langle\hat{\sigma}_{+}^{(b)}(\tau)\hat{\sigma}_{-}^{(c)}(\tau)\rangle.
\end{align}

\noindent Here, we have used Eq.~(\ref{eqs: A from S}) to obtain the second equation, employing also the spontaneous emission rate $\Gamma_{sp}$ (\ref{eqs: Sm sp}) and the radiation wavelength $\lambda$.

%

Hence, we can directly evaluate the intensity as soon as the correlations between atomic coherences are known.

\subsubsection*{Field correlation function and spectral intensity}

Another important field characteristic is its spectral intensity. According to the Wiener - Khinchin theorem, this quantity can be obtained from the Fourier transformation of the field correlation function. Using the same normalization as for the intensity (\ref{eqs: I za}) and expressing the field correlation function in terms of atomic variables, we find:

\begin{align}
\label{eqs: mean G za}
G^{(a)}(\tau_{1},\tau_{2}) & =\frac{2\epsilon_{0}\omega}{\Delta o\hbar c}\langle\hat{A}_{-}^{(a)}(\tau_{1})\hat{A}_{+}^{(a)}(\tau_{2})\rangle=\frac{3\Delta o}{32\pi\lambda^{2}}\Gamma_{sp}\sum_{b<a,c<a}\langle\hat{\sigma}_{+}^{(b)}(\tau_{1})\hat{\sigma}_{-}^{(c)}(\tau_{2})\rangle.
\end{align}

The two-time correlation function of atomic coherences $\langle\hat{\sigma}_{+}^{(b)}(\tau_{1})\hat{\sigma}_{-}^{(c)}(\tau_{2})\rangle$ cannot be related straightforwardly to the one-time correlation of atomic coherences (\ref{eqs: meaning of SpSm}) as obtained from the solution of (\ref{eqs: mean SpSm}). The quantum regression theorem, that is typically employed for establishing the connection between one-time and two-time correlations, requires the solution for one-atom averages \cite{1997'Scully}. In our case, the equations for one-atom averages that follow from (\ref{eqs: HL Sm}) are non-linear and the analytical solution is not straightforward.

In an alternative approach, we reformulate the problem of obtaining the field correlation function (\ref{eqs: mean G za}) into the problem of propagating it in space. For the difference between the field correlation function at position $z_a$ of atom $a$ and the position $z_{a+1}$ of neighboring atom $a+1$, we obtain the following expression within the approximation (\ref{eqs: approx triple factr}) (see details in Appendix \ref{appendix: Derivation of field correlation function propagation}):

\begin{align}
\label{eqs: mean dGdz}
&G^{(a+1)}(\tau_{1},\tau_{2})-G^{(a)}(\tau_{1},\tau_{2})  = \\ \nonumber
&\frac{3\Delta o}{8\pi}\Gamma_{sp} \left(\int\limits _{0}^{\tau_{1}}d\tau_{1}'e^{-(-i\Omega+\frac{\Gamma_{sp}}{2})(\tau_{1}-\tau_{1}')}\langle\hat{\sigma}_{z}^{(a)}(\tau_{1}')\rangle G^{(a)}(\tau_{1}',\tau_{2})+\int\limits _{0}^{\tau_{2}}d\tau_{2}'e^{-(i\Omega+\frac{\Gamma_{sp}}{2})(\tau_{2}-\tau_{2}')}\langle\hat{\sigma}_{z}^{(a)}(\tau_{2}')\rangle G^{(a)}(\tau_{1},\tau_{2}')\right)\\ \nonumber 
+&\frac{3\Delta o}{32\pi\lambda^{2}}\Gamma_{sp} \rho_{ee}^{(a)}(0)e^{i\Omega(\tau_{1}-\tau_{2})}e^{-\frac{\Gamma_{sp}}{2}(\tau_{1}+\tau_{2})}.
\end{align}

\subsection{Representation in continuous variables}

The set of equations (\ref{eqs: mean Sz}), (\ref{eqs: mean SpSm}), (\ref{eqs: mean dGdz}) forms a closed system of equations that enables us to derive the intensity and spectral properties of the emitted superfluorescence field. Since we are interested in macroscopic systems involving large numbers of atoms, it is advantageous to represent the obtained equations in terms of continuous quantities. With this aim, we introduce 

\begin{align}
\label{eqs: def continious}
\rho_{inv}(z,\tau)&:=\frac{1}{n\Delta z}\sum_{a:z<z_{a}<z+\Delta z}2\langle\hat{\sigma}_{z}^{(a)}(\tau)\rangle, \\ \nonumber
S(z_{1},z_{2},\tau)&:=\frac{1}{(n\Delta z)^{2}}\sum_{\substack{a:z_{1}<z_{a}<z_{1}+\Delta z \\ b:z_{2}<z_{b}<z_{2}+\Delta z}}\langle\hat{\sigma}_{+}^{(a)}(\tau)\hat{\sigma}_{-}^{(b)}(\tau)\rangle, \\ \nonumber
G(z,\tau_{1},\tau_{2})&:=\frac{1}{n\Delta z}\sum_{a:z<z_{a}<z+\Delta z}G^{(a)}(\tau_{1},\tau_{2}).
\end{align}

\noindent  Here, $n=n_v \pi R^2$ is the linear concentration of atoms ($n_v$ is volume concentration, $R$ is radius of cylindrical excited medium), $\rho_{inv}$ is the population inversion, $S$ is the correlation function of atomic coherences and $G$ is the correlation function of the field. As before, knowledge of the correlation function of atomic coherences is only required explicitly for different atoms $a\neq b$. In terms of these new variables, the system of equations (\ref{eqs: mean Sz}), (\ref{eqs: mean SpSm}) and (\ref{eqs: mean dGdz}) can be represented as the following system of integro-differential equations:

\begin{align}
\label{eqs: cont rinv}
\frac{\partial\rho_{inv}(z,\tau)}{\partial\tau}&=-2\Gamma_{sp}\rho_{ee}(z,\tau)-\frac{3\text{\ensuremath{\Delta o}}}{4\pi}\Gamma_{sp}n\int\limits _{0}^{z}dz'S(z,z',\tau),\\
\label{eqs: cont S}
\frac{\partial S(z_{1},z_{2},\tau)}{\partial\tau} & =-\Gamma_{sp}S(z_{1},z_{2},\tau)\\ \nonumber
+&\frac{3\text{\ensuremath{\Delta o}}}{16\pi}\Gamma_{sp}n_{1}\left(\rho_{inv}(z_{1},\tau)\int\limits _{0}^{z_{1}}dz_{1}'S(z_{1}',z_{2},\tau)+\rho_{inv}(z_{2},\tau)\int\limits _{0}^{z_{2}}dz_{2}'S(z_{1},z_{2}',\tau)\right)\\ \nonumber
+&\frac{3\text{\ensuremath{\Delta o}}}{16\pi}\Gamma_{sp}\left(\rho_{inv}(z_{1},\tau)\rho_{ee}(z_{2},\tau)\Theta(z_{1}-z_{2})+\rho_{inv}(z_{2},\tau)\rho_{ee}(z_{1},\tau)\Theta(z_{2}-z_{1})\right), \\ 
\label{eqs: cont G}
\frac{\partial G(z,\tau_{1},\tau_{2})}{\partial z} & = \\ \nonumber
&\frac{3\Delta o}{16\pi}\Gamma_{sp} n \left(\int\limits _{0}^{\tau_{1}}d\tau_{1}'e^{(i\Omega-\frac{\Gamma_{sp}}{2})(\tau_{1}-\tau_{1}')}\rho_{inv}(z,\tau_{1}')G(z,\tau_{1}',\tau_{2})\right. \\ \nonumber
&+\left.\int\limits _{0}^{\tau_{2}}d\tau_{2}'e^{-(i\Omega+\frac{\Gamma_{sp}}{2})(\tau_{2}-\tau_{2}')}\rho_{inv}(z,\tau_{2}') G(z,\tau_{1},\tau_{2}')\right)\\ \nonumber 
+&\frac{3\Delta o}{32\pi\lambda^{2}}\Gamma_{sp} n \rho_{ee}(z,0)e^{i\Omega(\tau_{1}-\tau_{2})}e^{-\frac{\Gamma_{sp}}{2}(\tau_{1}+\tau_{2})}.
\end{align}

\noindent Here, the occupation of excited states $\rho_{ee}(z,\tau)$ can be related to the population inversion as $\rho_{ee}(z,\tau)=[1+\rho_{inv}(z,\tau)]/2$  for a pure two-level system. Equations similar to (\ref{eqs: cont rinv}), (\ref{eqs: cont S}) were obtained in Ref.~\cite{1978'Ressayre} in the frame of the unrestricted Markovian master equation. Such an approach, however, is only capable to provide single-time observables. The two-time field correlation function, which is needed to investigate spectral properties, requires special treatment as in our case by equation (\ref{eqs: cont G}). Provided one is interested only in intensity profiles, then instead of solving (\ref{eqs: cont G}) one can also express the intensity directly in terms of solutions of (\ref{eqs: cont rinv}),(\ref{eqs: cont S}). Namely, one can use equation (\ref{eqs: I za}), separate terms with $b=c$ and express the result in terms of (\ref{eqs: def continious}):

\begin{align}
\label{eqs: cont I}
I(z,\tau) & =\frac{3\Delta o}{32\pi\lambda^{2}}\Gamma_{sp}\left(n^{2}\int\limits _{0}^{z}dz_{1}'\int\limits _{0}^{z}dz_{2}'S(z_{1}',z_{2}',\tau)+n\int\limits _{0}^{z}dz'\rho_{ee}(z',\tau)\right).
\end{align}


\section{\label{sec: Modification}Inclusion of pump process and incoherent processes}

\subsection{Incoherent processes relevant in  x-ray / XUV domain}

The system of equations (\ref{eqs: cont rinv}) - (\ref{eqs: cont G}) is capable of describing collective spontaneous emission from a medium that consists of pure two-level atoms. It also accounts for the swept instantaneous pumping of the atoms into their excited states. It is also possible to treat more general cases, where the excited state occupation resulting from the pump is position dependent through the use of any function $\rho_{ee}(z,0)$ as the initial condition. So far, however, we have always assumed that the condition $\rho_{ee}(z,t)+\rho_{gg}(z,t)=1$ is satisfied, which is insufficient for describing most of the experiments on collective spontaneous emission in x-ray / XUV domain. There, the levels,  between which the population inversion occurs, are populated and/or depleted as a result of multiple incoherent processes. Most significantly, this includes photoionization and can include Auger decay. Depending on the actual parameters of the system, the timescales of these processes can be comparable to the timescale of collective spontaneous emission. A strong point of our approach is that it can be generalized in a regular way to include processes of incoherent pumping, depletion and coherence decay, all of which affect the dynamics of the considered transition. 

\begin{figure}[tb]
\centerline{
  \resizebox{10cm}{!}{
  \begin{tikzpicture}[scale=0.75]
	\draw [line width=0.1cm, black] (2,1) -- (8,1);
	\node [below, black] at (5,1) {\ket{g}};
	\draw [line width=0.1cm, black] (2,3) -- (8,3);
	\node [above, black] at (5,3) {\ket{e}};
	\draw [line width=0.03cm, blue, ->] (0,2) -- (2.5,1);
	\node [above right,blue] at (1,1.5) {$r_g$};
	\draw [line width=0.07cm, blue, ->] (0,4) -- (2.5,3);
	\node [above right,blue] at (1,3.5) {$r_e$};
	\draw [line width=0.075cm, red, <->] (5,1) -- (5,3);
	\node [right,red] at (5,2) {$\Gamma_{sp}$};
	\draw [line width=0.05cm, brown, dashed, <->] (7,1) -- (7,3);
	\node [right,brown] at (7,2) {$q$};
	\draw [line width=0.05cm, gray, <-] (3,1) -- (3,3);
	\node [right,gray] at (3,2) {$\gamma_{n}$};
	\draw [line width=0.05cm,  gray, ->] (7.5,1) -- (10,2);
	\node [below,gray] at (9,1.5) {$\gamma_g$};
	\draw [line width=0.05cm,  gray, ->] (7.5,3) -- (10,4);
	\node [below,gray] at (9,3.5) {$\gamma_e$};
  \end{tikzpicture}
  }
}
\caption{Sketch of the two levels of interest subject to pumping with rates $r_e, r_g$, non-radiative decay with a rate $\gamma_{n}$, depletion with rates $\gamma_g, \gamma_e$, decoherence  with a rate $q$ and radiative transitions, which are treated self-consistently within our approach -- the typical value determining the evolution rate being $\Gamma_{sp}$.}
\label{fig: level scheme}
\end{figure}
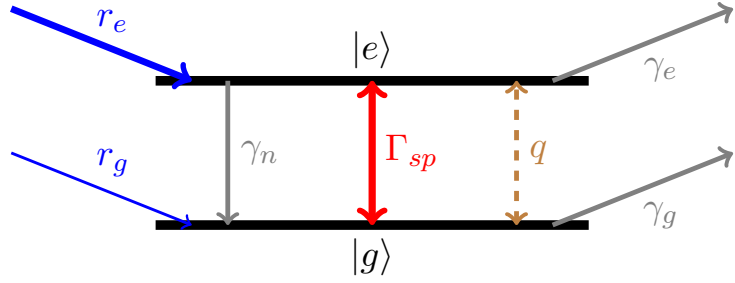

In order to provide a formalism that can be coupled to further rate equations describing the evolution of atoms in the field of a strong XFEL pulse (cf. rate equation codes like Ref.~\cite{2016'Jurek}), we consider two levels of interest (\ket{e},\ket{g}) under the influence of several incoherent processes. A general scenario is outlined in Fig.~\ref{fig: level scheme}, including the following effects: i) pumping of the ground and excited levels with rates $r_g, r_e$. Examples of pumping include photoionization by an XFEL pulse \cite{2012'Rohringer}, \cite{2015'Yoneda}, \cite{2018'Kroll}, \cite{2014'Weninger} and Auger decay \cite{1975'McGuire}, \cite{1986'Kapteyn}, \cite{1988'Kapteyn}; ii) non-radiative transitions from \ket{e} to \ket{g} with a rate $\gamma_{n}$; iii) decoherence between levels \ket{e}, \ket{g} with a rate $q$, as caused -- for example -- by collisions \cite{1986'Kapteyn}, \cite{1988'Kapteyn}; and iv) depletion of the levels with rates $\gamma_g, \gamma_e$, e.g., through further photoionization by the XFEL pump or decay processes like Auger decay; besides these processes, radiative decay from \ket{e} to \ket{g} and induced transitions between \ket{e} and \ket{g} are already modeled within our current approach. The quantity that determines the coupling between the levels and the emitted radiation is $\Gamma_{sp}$. Note that the rates $r_g, r_e, q,\gamma_g, \gamma_e$ may be conditioned by the external XFEL pulse or the local atomic density, hence they can depend on position and time explicitly.

\subsection{Description of incoherent processes in terms of Lindblad superoperators}

Incoherent processes are modeled as an interaction of the system with an appropriate reservoir. We assume that the reservoirs for all processes and for all atoms are independent and furthermore that these incoherent processes can be described in the Markovian approximation. In this way, we can eliminate the reservoir's variables and describe the transitions between the levels of interest with the help of a master equation \cite{2007'Meystre}, \cite{1993'Briegel}

\begin{align}
\label{eqs: me general}
\frac{\partial\hat{\rho}^{(a)}}{\partial t}=\frac{1}{i\hbar}\left[\hat{H},\hat{\rho}^{(a)}\right]+L^{(a)}\{\hat{\rho}^{(a)}\}.
\end{align}

\noindent Here, $L^{(a)}\{\hat{\rho}^{(a)}\}$ denotes the non-unitary part of the evolution, which describes transitions due to  the incoherent process of interest. It can be considered as a Lindblad superoperator acting on the density matrix. For the processes depicted in Fig. \ref{fig: level scheme} we use Lindblad superoperators of the following form \cite{1993'Briegel}:

\begin{align}
\label{eqs: def Lrl}
L_{r,l}^{(a)}\{\hat{\rho}^{(a)}\}&=\frac{1}{2}\tilde{r}_l(z_a,\tau)\left(2\hat{\sigma}_{+,xl}^{(a)}\hat{\rho}^{(a)}\hat{\sigma}_{-,xl}^{(a)}-\hat{\sigma}_{-,xl}^{(a)}\hat{\sigma}_{+,xl}^{(a)}\hat{\rho}^{(a)}-\hat{\rho}^{(a)}\hat{\sigma}_{-,xl}^{(a)}\hat{\sigma}_{+,xl}^{(a)}\right), \\
\label{eqs: def Lg}
L_{\gamma,l}^{(a)}\{\hat{\rho}^{(a)}\}&=\frac{1}{2}\gamma_l(z_a,\tau)\left(2\hat{\sigma}_{-,xl}^{(a)}\hat{\rho}^{(a)}\hat{\sigma}_{+,xl}^{(a)}-\hat{\sigma}_{+,xl}^{(a)}\hat{\sigma}_{-,xl}^{(a)}\hat{\rho}^{(a)}-\hat{\rho}^{(a)}\hat{\sigma}_{+,xl}^{(a)}\hat{\sigma}_{-,xl}^{(a)}\right), \\
\label{eqs: def Ln}
L_{n}^{(a)}\{\hat{\rho}^{(a)}\}&=\frac{1}{2}\gamma_n\left(2\hat{\sigma}_{-}^{(a)}\hat{\rho}^{(a)}\hat{\sigma}_{+}^{(a)}-\hat{\sigma}_{+}^{(a)}\hat{\sigma}_{-}^{(a)}\hat{\rho}^{(a)}-\hat{\rho}^{(a)}\hat{\sigma}_{+}^{(a)}\hat{\sigma}_{-}^{(a)}\right), \\
\label{eqs: def Lq}
L_{q}^{(a)}\{\hat{\rho}^{(a)}\}&=\frac{1}{4}q(z_a,\tau)\left(4\hat{\sigma}_{z}^{(a)}\hat{\rho}^{(a)}\hat{\sigma}_{z}^{(a)}-\hat{\rho}^{(a)}\right).
\end{align} 


\noindent With $l=e,g$, (\ref{eqs: def Lrl}) corresponds to pumping and (\ref{eqs: def Lg}) to depletion of level \ket{l}, $\tilde{r}_l$ denotes the pumping rate normalized to the occupation of the state, from which pumping takes place (see details in Appendix \ref{appendix: Modification of equations for atomic and field observables due to incoherent processes}). Analogously, $\gamma_l$ is the depletion rate. Non-radiative decay with a rate $\gamma_n$ is described by (\ref{eqs: def Ln}) and decoherence with rate $q$ is described by (\ref{eqs: def Lq}). The operators $\hat{\sigma}_{\pm,xl}^{(a)}$ account for transitions of atom $a$ between an auxiliary level \ket{x} and level \ket{l}:

\begin{align}
\label{eqs: def Sxl}
\hat{\sigma}_{+,xl} = |l \rangle \langle x |, \hat{\sigma}_{-,xl} = |x \rangle \langle l |.
\end{align}

\noindent Finally, the total Lindblad superoperator in (\ref{eqs: me general}) becomes:

\begin{align}
\label{eqs: L tot}
L^{(a)}=L_{r,e}^{(a)}+L_{r,g}^{(a)}+L_{n}^{(a)}+L_{q}^{(a)}+L_{\gamma,e}^{(a)}+L_{\gamma,g}^{(a)}.
\end{align}

\noindent The modification of equations (\ref{eqs: mean Sz}), (\ref{eqs: mean SpSm}), (\ref{eqs: mean dGdz}) due to (\ref{eqs: L tot}) can be derived with the help of quantum Einstein relations \cite{1997'Scully} (see the derivation in Appendix \ref{appendix: Modification of equations for atomic and field observables due to incoherent processes} for details). In section \ref{sec: sub: Modified correlation function propagation equations}, we give a summary of the ensuing changes. 

\subsection{Absorption of the emitted field}

Another important process that should be taken into account for a realistic treatment of collective spontaneous emission in x-ray / XUV domain is non-resonant absorption of the emitted radiation. It can be accounted for by adding an imaginary part to the field wavevector: $\vec k \rightarrow \vec k+ i \frac{\kappa}{2} \hat{\vec k}$, where $\kappa$ is absorption coefficient. Incorporation of this factor in (\ref{eqs: H-L general A}) after performing the 1D approximation modifies (\ref{eqs: A from S}) to

\begin{align}
\label{eqs: abs A from S}
\hat{A}_{+}^{(a)}(\tau) & =-\frac{i \Delta o e p^* \Omega}{16\pi^{2}\epsilon_{0} m c^{2}}e^{-\frac{\kappa}{2}(z_a-z_b)}\sum_{b<a}\hat{\sigma}_{-}^{(b)}(\tau).
\end{align}

\subsection{Equations of motion for atomic system and field radiation}
\label{sec: sub: Modified correlation function propagation equations}

Collecting the results from Appendix \ref{appendix: Modification of equations for atomic and field observables due to incoherent processes} and adding the absorption factor that appeared in (\ref{eqs: abs A from S}) we arrive at the following system of equations:

\begin{align}
\label{eqs: cont LA ee}
\frac{\partial\rho_{ee}(z,\tau)}{\partial\tau}&=r_e(z,\tau)-\Gamma_{ee}(z,\tau)\rho_{ee}(z,\tau)-\frac{3\text{\ensuremath{\Delta o}}}{8\pi}\Gamma_{sp}n\int\limits _{0}^{z}dz'\mathcal{A}(z,z')S(z,z',\tau),\\ 
\label{eqs: cont LA gg}
\frac{\partial\rho_{gg}(z,\tau)}{\partial\tau}&=r_g(z,\tau)+(\Gamma_{sp}+\gamma_{n})\rho_{ee}(z,\tau)-\gamma_{g}(z,\tau)\rho_{gg}(z,\tau)\\ \nonumber
&+\frac{3\text{\ensuremath{\Delta o}}}{8\pi}\Gamma_{sp}n\int\limits _{0}^{z}dz'\mathcal{A}(z,z')S(z,z',\tau),\\
\label{eqs: cont LA S}
\frac{\partial S(z_{1},z_{2},\tau)}{\partial\tau} & = -\frac{1}{2}\left[\Gamma(z_1,\tau)+\Gamma(z_2,\tau)\right]S(z_{1},z_{2},\tau)\\ \nonumber
+&\frac{3\text{\ensuremath{\Delta o}}}{16\pi}\Gamma_{sp}n\left[\rho_{inv}(z_{1},\tau)\int\limits _{0}^{z_{1}}dz_{1}'\mathcal{A}(z_1,z_1')S(z_{1}',z_{2},\tau)\right.\\ \nonumber
+&\left.\rho_{inv}(z_{2},\tau)\int\limits _{0}^{z_{2}}dz_{2}'\mathcal{A}(z_2,z_2')S(z_{1},z_{2}',\tau)\right]\\ \nonumber
+&\frac{3\text{\ensuremath{\Delta o}}}{16\pi}\Gamma_{sp}\left[\rho_{inv}(z_{1},\tau)\rho_{ee}(z_{2},\tau)\mathcal{A}(z_1,z_2)\Theta(z_{1}-z_{2})\right.\\ \nonumber
+&\left.\rho_{inv}(z_{2},\tau)\rho_{ee}(z_{1},\tau)\mathcal{A}(z_2,z_1)\Theta(z_{2}-z_{1})\right], \\ 
\label{eqs: cont LA G}
\frac{\partial G(z,\tau_{1},\tau_{2})}{\partial z} & = -\kappa(z) G(z,\tau_{1},\tau_{2}) \\ \nonumber
&+\frac{3\Delta o}{16\pi}\Gamma_{sp} n \left[\int\limits _{0}^{\tau_{1}}d\tau_{1}'e^{i\Omega(\tau_{1}-\tau_{1}')}\mathcal{D}(z,\tau_1,\tau_1')\rho_{inv}(z,\tau_{1}')G(z,\tau_{1}',\tau_{2})\right. \\ \nonumber
&+\left.\int\limits _{0}^{\tau_{2}}d\tau_{2}'e^{-i\Omega(\tau_{2}-\tau_{2}')}\mathcal{D}(z,\tau_2,\tau_2')\rho_{inv}(z,\tau_{2}') G(z,\tau_{1},\tau_{2}')\right]\\ \nonumber 
+&\frac{3\Delta o}{32\pi\lambda^{2}}\Gamma_{sp} n e^{i\Omega(\tau_{1}-\tau_{2})}\left\lbrace\vphantom{\int\limits _{0}^{\min{\tau_1,\tau_2}}}\mathcal{D}(z,\tau_1,0)\mathcal{D}(z,\tau_2,0)\rho_{ee}(z,0)\right.   \\ \nonumber
&+\left.\int\limits _{0}^{\min{\tau_1,\tau_2}}d\tau'\mathcal{D}(z,\tau_1,\tau')\mathcal{D}(z,\tau_2,\tau')
\left[r_e(z,\tau')+\left(\Gamma(z,\tau')-\Gamma_{ee}(z,\tau')\right)\rho_{ee}(z,\tau')\right]\right\rbrace.
\end{align}

\noindent Here, decay rates $\Gamma(z,\tau'),\Gamma_{ee}(z,\tau')$ have been introduced (see Appendix \ref{appendix: Modification of equations for atomic and field observables due to incoherent processes} for derivations) and read

\begin{align}
\label{eqs: def Gtotal Gammaee}
\Gamma(z,\tau)&=\Gamma_{sp}+\gamma_n+q(z,\tau)+\gamma_e(z,\tau)+\gamma_{g}(z,\tau), \\ \nonumber
\Gamma_{ee}(z,\tau)&=\Gamma_{sp}+\gamma_e(z,\tau)+\gamma_n.
\end{align}

\noindent The term $\mathcal{A}$ is responsible for field absorption and is assumed to be position dependent; the term $\mathcal{D}$ is responsible for decoherence due to spontaneous decay and other incoherent processes. The explicit expressions read:

\begin{align}
\label{eqs: damping terms A D}
\mathcal{A}(z_2,z_1)=e^{-\frac{1}{2}\int_{z_1}^{z_2}dz'\kappa(z')}, \quad \mathcal{D}(z,\tau_2,\tau_1)=e^{-\frac{1}{2}\int_{\tau_1}^{\tau_2}d\tau'\Gamma(z,\tau')}.
\end{align} 

The intensity of the emitted field expressed in terms of atomic variables is modified to

\begin{align}
\label{eqs: cont LA I}
I(z,\tau) & =\frac{3\Delta o}{32\pi\lambda^{2}}\Gamma_{sp}\left(n^{2}\int\limits _{0}^{z}dz_{1}'\int\limits _{0}^{z}dz_{2}'\mathcal{A}(z,z_1')\mathcal{A}(z,z_2')S(z_{1}',z_{2}',\tau)+n\int\limits _{0}^{z}dz'\mathcal{A}^2(z,z')\rho_{ee}(z',\tau)\right).
\end{align}

\section{\label{sec: Numerical examples} Discussion and numerical examples}

The system of equations (\ref{eqs: cont LA ee}) - (\ref{eqs: cont LA G}) provides a unified treatment of incoherent and coherent processes taking place during the collective spontaneous emission.  Namely, as one limiting case it allows for the description of incoherent spontaneous emission by individual atoms. In the opposite limit it includes the propagation of coherent electromagnetic pulses in a medium of two-level atoms, which can be reformulated in the terms of semi-classical Maxwell-Bloch equations \cite{1997'Scully}. Let us consider each of limiting cases in more detail.

\subsection{Spontaneous emission} 

An important limiting case that should be reproduced by our approach is the number of spontaneously emitted photons per unit time into the solid angle $\Delta o$. A reference for this quantity can be calculated based on the usual quantum optical formalism, see Refs.~\citep{1974'Agarwal}, \cite{1997'Scully}. This employs the expression of the emitted field in terms of atomic transition operators, see \citep{1997'Scully} Ch.10A, and results in

\begin{align}
\label{eqs: Nsp exact}
\frac{d }{d\tau}N_{\text{ph.,sp.}}(\tau, \theta=\pi/2)=\frac{2 \epsilon_0 c}{\hbar \omega}\langle \hat{\vec{E}}_{-}(\tau, \theta=\frac{\pi}{2})\hat{\vec{E}}_{+}(\tau, \theta=\frac{\pi}{2}) \rangle n z r^2 \Delta o= \frac{3}{8\pi}\Gamma_{sp} \rho_{ee}(\tau) n z \Delta o.
\end{align}

\noindent Here, we have used the value of the spontaneous emission rate (\ref{eqs: Gsp def}); $r$ is the distance from the atoms to a remote observation point; $\theta$ is the angle between the emission direction and the vector of the atomic dipole moment, which we consider to be $\theta=\pi/2$ for forward emission; $z$ is the length of the system.

Within our approach, i.e., Eqs.~(\ref{eqs: cont LA ee}) - (\ref{eqs: cont LA G}) one can obtain spontaneous emission rate if one neglects correlation between atomic coherences. Then from (\ref{eqs: cont LA I}) according to the used convention for intensity (\ref{eqs: I za}) one obtains

\begin{align}
\label{eqs: Nsp our}
\frac{d N_{\text{ph.,sp.}}(\tau)}{d\tau}&=I(z,\tau) \Delta o \pi R^2 =\frac{3}{8\pi}\Gamma_{sp}\rho_{ee}(\tau)nz\Delta o \cdot \xi,\\
\xi&=\frac{\Delta o \pi R^2}{4\lambda^{2}}.
\end{align}

\noindent This expression is in agreement with the exact result of Eq. (\ref{eqs: Nsp exact}) up to a factor $\xi$ that appears due to the 1D approximation (\ref{eqs: 1D approximation}). The actual value of $\Delta o$ is subject to the detail of the approximation, however, the typical estimate based on diffraction $\Delta o \sim \lambda^2 / R^2$ is in agreement with $\xi \sim 1$. 

The spectral properties could be obtained directly from the expression for the two-time correlation function (\ref{eqs: cont G}). In the case of spontaneous emission one can drop the term involving the field correlation function on the r.h.s. of (\ref{eqs: cont G}) and obtain the time dependence of the form $e^{i\Omega(\tau_{1}-\tau_{2})}e^{-\frac{\Gamma_{sp}}{2}(\tau_{1}+\tau_{2})}$. This form agrees with the exact result for a single two-level atom, see e.g. \cite{1974'Agarwal} Eq.(10.14b), and results in a Lorentzian spectrum.

\subsection{Maxwell - Bloch equations} 

Under the conditions that a large number of photons has been emitted and essential correlations between atomic coherences have been built up, one can expect that the behavior of the system can be described semi-classically. This is typically done by means of Maxwell-Bloch equations, which are widely used for the description of superradiance in the optical domain \citep{1982'Gross}, \citep{2017'Kocharovsky} as well as for the description of collective spontaneous emission in the x-ray / XUV domain \citep{2000'Larroche}, \citep{2014'Weninger}. We shall subsequently outline the connection between our correlation function approach (\ref{eqs: cont LA ee}) - (\ref{eqs: cont LA G}) and Maxwell-Bloch equations.

A crude way to obtain semi-classical equations is to replace operators by $c$-numbers in the Heisenberg equations of motion. If one carried out this replacement in (\ref{eqs: A from S})-(\ref{eqs: HL Sz}), one would obtain equations that are similar to (\ref{eqs: cont LA ee}) - (\ref{eqs: cont LA G}) except for the last term in both (\ref{eqs: cont LA S}) and (\ref{eqs: cont LA G}). These last terms are responsible for spontaneous emission. During the derivation of them, we have essentially used quantum properties of operators (\ref{eqs: def Spm}) --  in particular $\langle\hat{\sigma}_{+}(\tau)\hat{\sigma}_{-}(\tau)\rangle=\rho_{ee}(\tau)$. This property is crucial for the correct description of spontaneous emission and approaches based on the semi-classical relation $\langle\hat{\sigma}_{+}(\tau)\hat{\sigma}_{-}(\tau)\rangle=\sigma_{+}(\tau)\sigma_{-}(\tau)$ (sometimes referred to as neoclassical theories) cannot reproduce the associated exponential decay \citep{1974'Agarwal}.

If the dynamics is dominated by the emitted field, we can, however, neglect the spontaneous terms in (\ref{eqs: cont LA ee}) - (\ref{eqs: cont LA G}) safely. Using this approximation, (\ref{eqs: cont LA ee}) - (\ref{eqs: cont LA G}) reduce to Maxwell-Bloch equations, if we also assume the correlation functions to factorize:

\begin{align}
\label{eqs: MB factorization S}
S(z_1,z_2,\tau)&=\rho_{ge}(z_1,\tau)\rho_{eg}(z_2,\tau), \\ 
\label{eqs: MB factorization G}
G(z,\tau_1,\tau_2)&=\frac{2\epsilon_{0} c}{\Delta o \hbar \Omega}e^{i\Omega(\tau_{1}-\tau_{2})}\mathcal{E}_{-}(z,\tau_{1})\mathcal{E}_{+}(z,\tau_{2}).
\end{align}

\noindent Here, $\rho_{ge},\rho_{eg}$ are off-diagonal elements of the density matrix (coherences). In (\ref{eqs: MB factorization G}), we have expressed the field correlation function in terms of the slowly varying electric field envelopes $\mathcal{E}_{\pm}(z,\tau)$ in order to be compliant with typical Maxwell-Bloch formulations \cite{1982'Gross}, \cite{2000'Larroche}. Alltogether, we obtain from (\ref{eqs: cont LA ee}) - (\ref{eqs: cont LA G}) equations for the density matrix elements and the field (see details in Appendix \ref{appendix: Derivation of Maxwell-Bloch equations from correlation fucntion equations}):

\begin{align}
\label{eqs: MB ee}
\frac{\partial\rho_{ee}(z,\tau)}{\partial\tau}&=r_e(z,\tau)-\Gamma_{ee}(z,\tau)\rho_{ee}(z,\tau)-\frac{2 i \mu}{\hbar}\mathcal{E}_{-}(z,\tau)\rho_{eg}(z,\tau),\\ 
\label{eqs: MB gg}
\frac{\partial\rho_{gg}(z,\tau)}{\partial\tau}&=r_g(z,\tau)+(\Gamma_{sp}+\gamma_{n})\rho_{ee}(z,\tau)-\gamma_{g}(z,\tau)\rho_{gg}(z,\tau)-\frac{2 i \mu}{\hbar}\mathcal{E}_{+}(z,\tau)\rho_{ge}(z,\tau),\\
\label{eqs: MB ge}
\frac{\partial \rho_{ge}(z,\tau)}{\partial\tau}&=-\frac{\Gamma(z,\tau)}{2}\rho_{ge}(z,\tau)+\frac{i \mu}{\hbar}\rho_{inv}(z,\tau)\mathcal{E}_{-}(z,\tau),\\
\label{eqs: MB E}
\frac{\partial \mathcal{E}_{+}(z,\tau)}{\partial z} &=-\frac{\kappa(z)}{2}\mathcal{E}_{+}(z,\tau) + \frac{i \Omega}{2 \epsilon_0 c}\mu n_3 \rho_{eg}(z,\tau).
\end{align}

This system of equations agrees with the typically used Maxwell-Bloch system of equations \cite{1982'Gross}. As stated before, however, it shows essential weakness in the treatment of collective spontaneous emission: assuming the initial absence of fields and coherences, neither of them will appear later on according to (\ref{eqs: MB ge}), (\ref{eqs: MB E}). This is an immediate consequence of neglecting the spontaneous emission terms in the transition from (\ref{eqs: cont LA ee}) - (\ref{eqs: cont LA G}) to (\ref{eqs: MB ee}) - (\ref{eqs: MB E}). There are a number of semi-phenomenological ways of re-introducing spontaneous emission into the Maxwell-Bloch equations (\ref{eqs: MB ee}) - (\ref{eqs: MB E}). For applications connected with spontaneous emission in x-ray / XUV domain, the most widespread approach is to add random noise term on the r.h.s. of the equations for the coherences \citep{2000'Larroche}, \cite{2007'Almiev}, \citep{2010'Kim}, \citep{2014'Wang}, \citep{2014'Weninger}, \citep{2015'Depresseux}, \cite{2018'Kroll}, \cite{2018'Lyu}. In the frame of this approach, the noise terms are assumed to have the following properties

\begin{align}
\label{eqs: s eq1}
\langle s_{\pm}(z,t) \rangle =0, \quad \langle s_{+}(z_1,t_1) s_{-}(z_2,t_2) \rangle =\rho_{ee}(z_1,t_1) F \delta(z_1-z_2)\delta(t_1-t_2),
\end{align}

\noindent where $\langle ... \rangle$ means statistical average over multiple realizations. The coefficient $F$ is chosen to fit the value of the spontaneous emission. This and the resulting temporal profile of the spontaneous emission were recently discussed in detail in Ref.~\cite{2018'Krusic}. As it was shown there, this kind of noise model does not reproduce the expected exponential decay. As an example, we consider simplified case without incoherent processes and complete initial excitation in Appendix \ref{appendix: Estimation of noise-term correlation factor for Maxwell-Bloch equations}. For this scenario, the Maxwell-Bloch equations with noise terms (\ref{eqs: s eq1}) predict a temporal behavior as $\Gamma_{sp}t\, e^{-\Gamma_{sp} t}$, which can be tolerable for cases with a short decoherence time. However, for any choice of parameters, the Maxwell-Bloch equations with noise terms (\ref{eqs: s eq1}) will produce an unrealistic, time-delayed peak of the spontaneous emission.

\begin{figure}
\includegraphics[width=1\linewidth, scale=1]{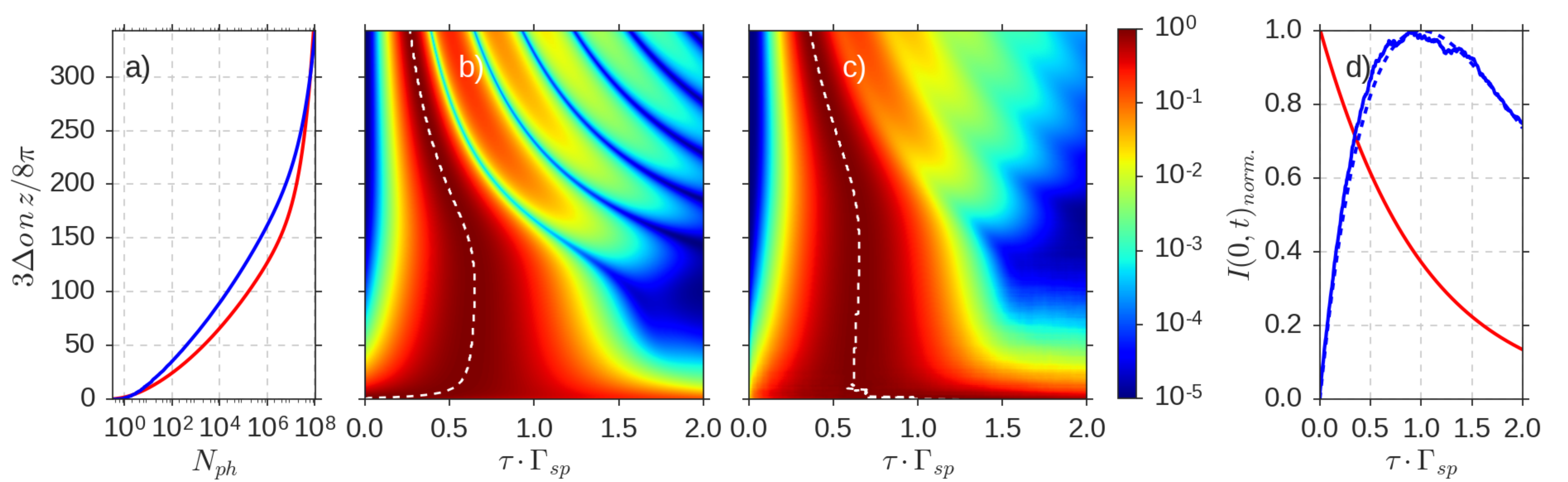}%
\caption{\label{fig: fig_CF_vs_MB}  a - simulation of emitted photon number versus system length (red line - correlation function calculation, blue line - Maxwell-Bloch approach),  b - simulation of temporal intensity profiles for systems of different length $z$, calculation according to the correlation function equations (\ref{eqs: cont rinv})-(\ref{eqs: cont S}) and (\ref{eqs: cont I}), dashed white line shows the position of the intensity maximum; c - the same figure obtained from the Maxwell-Bloch equations (\ref{eqs: MB ee}) - (\ref{eqs: MB E}) with noise terms (\ref{eqs: s eq1}) and (\ref{eqs: noise eq5 F}), averaged over 100 realizations; d - section of the same data taken at $z=0$, the correlation function calculation (red curve) coincides with $e^{-\Gamma_{sp}\tau}$, while the Maxwell-Bloch result averaged over 1000 realizations (blue curve) approaches $\Gamma_{sp}\tau e^{-\Gamma_{sp}\tau} e$ (dashed blue line).}
\end{figure}

A visual comparison of the temporal intensity profiles based on the Maxwell-Bloch equations (using the noise term correlation factor derived in Appendix \ref{appendix: Estimation of noise-term correlation factor for Maxwell-Bloch equations}) with calculations according to our correlation function equations (\ref{eqs: cont rinv})-(\ref{eqs: cont S}), (\ref{eqs: cont I}) is given in Fig.\ref{fig: fig_CF_vs_MB}. Here, a case of instantaneous swept pumping is considered, namely, a complete population inversion is assumed at $\tau=0$. In Fig.\ref{fig: fig_CF_vs_MB} b,c, the radiated intensity is plotted versus retarded the time $\tau$ and the length of the medium $z$. For each value of $z$, the intensity is normalized to the maximal value. Fig.\ref{fig: fig_CF_vs_MB} b shows results based on (\ref{eqs: cont rinv})-(\ref{eqs: cont S}), (\ref{eqs: cont I}), whereas Fig.\ref{fig: fig_CF_vs_MB} c is obtained from the Maxwell-Bloch approach. From (\ref{eqs: cont rinv})-(\ref{eqs: cont S}), (\ref{eqs: cont I}), one can see that if one uses the dimensionless time $\tau \Gamma_{sp}$ and the dimensionless length $3\Delta o n z/8\pi$ then the only parameter that defines the temporal shape of the intensity profile of the system is the solid angle $\Delta o$. The value $\Delta o=4\cdot 10^{-6}$ was used for the simulations presented in Fig.\ref{fig: fig_CF_vs_MB}.  

At short length of the medium, the stimulated emission can be neglected, hence the temporal evolution of the intensity should have the form of spontaneous decay $e^{-\Gamma_{sp}\tau}$. Fig.\ref{fig: fig_CF_vs_MB} d shows sections of Fig.\ref{fig: fig_CF_vs_MB} b,c close to $z=0$. As we discussed above, our approach shows the expected exponential decay behavior, while the Maxwell-Bloch approach follows a time dependence like $\Gamma_{sp}t\, e^{-\Gamma_{sp} t}$. The calculated temporal profile shows noisy behavior even after averaging over 1000 realizations. With increasing system length $z$, the stimulated emission results in exponential-like growth of the emitted photons number, see Fig.\ref{fig: fig_CF_vs_MB} a. During this process, the temporal intensity profile transforms from exponential decay (peaked at $\tau=0$) to a profile showing a delayed peak, see Fig.\ref{fig: fig_CF_vs_MB} b. In the frame of the Maxwell-Bloch approach, the temporal profile always shows a delayed peak, even at infinitesimally small $z$. The position of this peak is strongly influenced by noise at small $z$ and stays approximately constant up to the onset of saturation, see Fig.\ref{fig: fig_CF_vs_MB} c. After the saturation sets in (after $3\Delta o n z/8\pi \sim 150$), oscillatory behavior (ringing) appears. These oscillations are smeared in the Maxwell-Bloch approach due to the finite number of realizations. In both models, the period of the oscillations decreases with increasing system length $z$ and the position of the peak tends to shorter times.

\subsection{Numerical examples} 
\begin{figure}[tb]
\includegraphics[width=1\linewidth, scale=1]{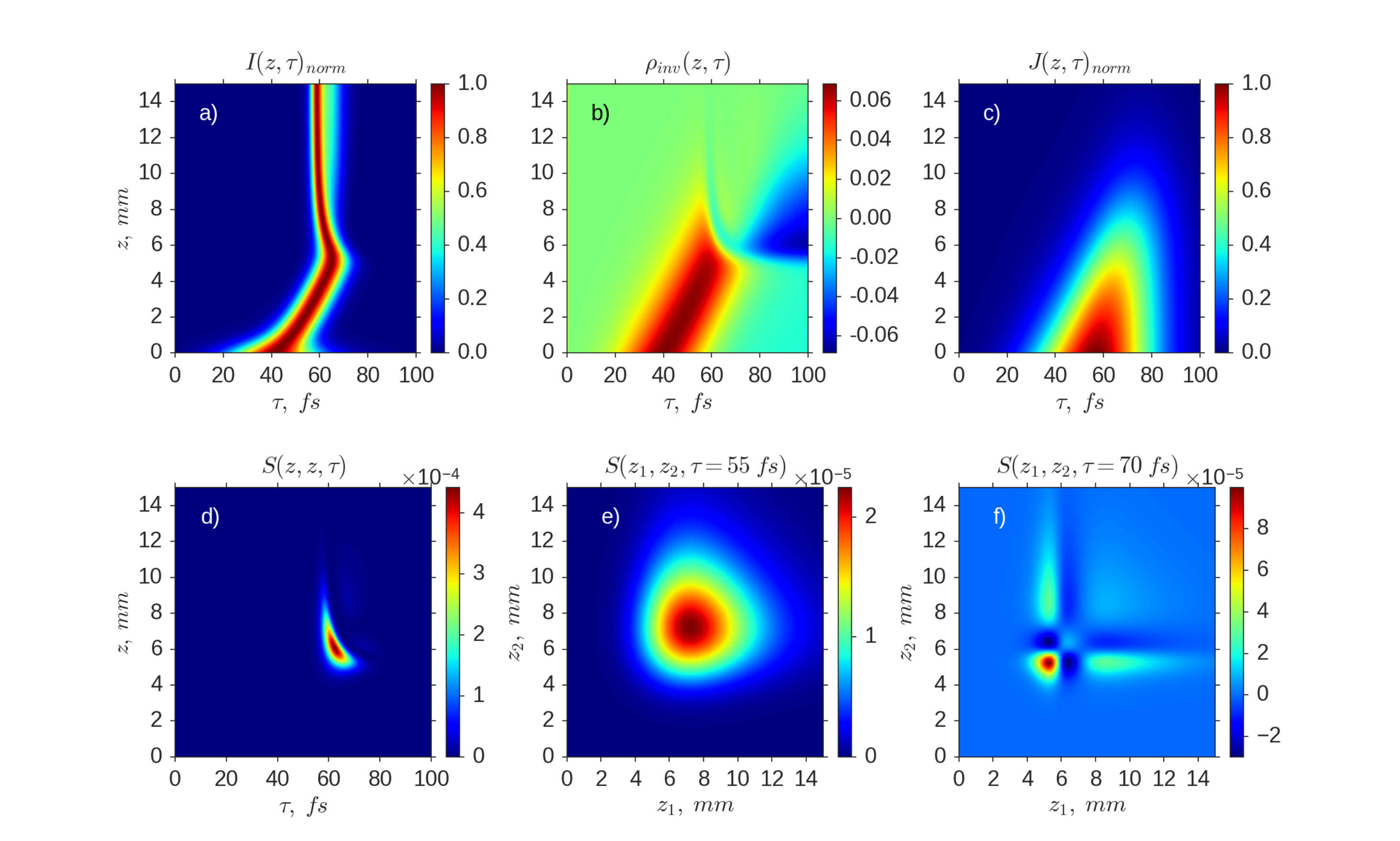}%
\caption{\label{fig: fig_CF_Ne_example} Evolution of collective spontaneous emission induced by an XFEL pump in Ne gas, a - profile of emitted superfluorescence intensity normalized to 1 for each $z$ value, b - population inversion, c - profile of the XFEL pump pulse normalized to maximal value, d - diagonal part of correlation function of atomic coherences, e,f - values of correlation function of atomic coherences taken at time moments 55 fs and 70 fs.}
\end{figure}

The correlation function equations (\ref{eqs: cont LA ee}) - (\ref{eqs: cont LA G}) can be used to model various realistic systems, provided the 1D approximation is applicable. As a first example let us consider evolution induced by an XFEL pulse in Neon gas. Calculations for Fig. \ref{fig: fig_CF_Ne_example} were done at parameters corresponding to experiment described in \citep{2012'Rohringer} and aim to compare with Maxwell-Bloch calculations performed for the same system in paper \cite{2014'Weninger}. In particular, the Neon gas cell with density 1.6 $\cdot 10^{19} $ cm$^{-3}$ and length 15 mm is considered to be irradiated by a an XFEL pulse with photon energy 880 eV, containing $2 \cdot 10^{12} $ photons within 40 fs FWHM Gaussian temporal profile, focusing with 2 $\mu$m radius is assumed. The photoionization of $1s$ electrons with crossection 0.3 Mb results in population inversion between $1s^1 2s^2 2p^6$ and $1s^2 2s^2 2p^5$ states. The transition between these states results in emission at 1.46 nm wavelength, the spontaneous lifetime is 160 fs, the competing process is Auger decay with 2.4 fs decay time. 

In Fig. \ref{fig: fig_CF_Ne_example} a the temporal intensity profile normalized at each value of $z$ (similar to Fig.\ref{fig: fig_CF_vs_MB} b,c) is shown. The temporal position of the peak at small $z$ values is determined by excitation of atoms. The excitation is conditioned by the pump pulse, its spatio-temporal evolution due to absorption is shown in Fig.\ref{fig: fig_CF_Ne_example} c. Fig.\ref{fig: fig_CF_Ne_example} b shows the population inversion. Before saturation (at $\sim 6$ mm) the evolution of population inversion is mainly determined by the pump, the peak of intensity comes shortly after the peak of population inversion. After saturation, oscillatory behavior can be seen in the population inversion, the peak of intensity temporal profile shifts to shorter times. The evolution of observables presented in Fig. \ref{fig: fig_CF_Ne_example} a-c can be compared directly to that in Fig.5 a-c in paper \cite{2014'Weninger}. To have qualitative agreement for propagation distances above $\sim$2 mm the value of solid angle $\Delta o$ was taken as $2\pi R^2/L^2$. Despite of qualitative agreement for larger propagation distances, the Maxwell-Bloch calculation of intensity profile below $\sim$2 mm of propagation show fluctuating behavior due to stochastic noise terms. As we have seen in Fig.\ref{fig: fig_CF_vs_MB}, the Maxwell-Bloch approach with noise term does not describe the initial amplification stage adequately. The analytical treatment of this stage for the considered system is complicated due to pumping and Auger processes taking place on the same timescale as the emission. Therefore, the application of analytical Bessel-function based solutions known for instantaneously pumped systems \cite{1979'Haake}, \cite{1979'Polder}, \cite{1982'Gross} is not possible.  Under these conditions, the presented approach is the method of choice to trace the evolution from spontaneous emission to amplified emission and superfluorescence.

The correlation function of atomic coherences $S(z_1,z_2,\tau)$ can be considered as source of stimulated emission, see Eq. (\ref{eqs: cont LA I}). Several sections of this function are shown in Fig. \ref{fig: fig_CF_Ne_example} d-f. Specifically, Fig. \ref{fig: fig_CF_Ne_example} d shows the correlation function of the atomic coherence at infinitesimally close points $z_1=z_2=z$ at retarded time moments $\tau$. Its peak at $\sim 60$ fs, $\sim 6 $ mm corresponds to the onset of oscillations in the population inversion (see Fig. \ref{fig: fig_CF_Ne_example} b) and saturation. The two-point correlation function $S(z_1,z_2,\tau)$ for a fixed time moment $\tau$ represents purely spatial correlation of the polarization of the medium, and is represented for times $\tau=55$ fs and $\tau=70$ fs in Fig. \ref{fig: fig_CF_Ne_example} e,f, i.e. for times before and after the maximum of $S(z,z,\tau)$.  Before onset of saturation ($\tau=55$ fs) the correlation function $S(z,z,\tau)$ shows a single maximum. After saturation, several local maxima and minima (anti-correlation) are visible. Beyond saturation, the population of the excited state drops below the population of the ground state (negative population inversion) in some parts of the medium. This results in the absorption of the emitted field and corresponds to negative values of the correlation function of atomic coherences, see Fig. \ref{fig: fig_CF_Ne_example} f, while other parts of the sample show positive correlation (emission). This partitioning of the sample into positively and negatively correlated parts is a direct manifestation of the temporal ringing in the spatial domain.

As another example, we consider collective spontaneous emission from Xe atoms pumped by a XFEL pulse at a mean photon energy of 73 eV corresponding to conditions of a recent experiment that was conducted at the soft x-ray free-electron laser FLASH \cite{2018'Mercadier}. Namely, following parameters were used: pump photon energy is 73 eV, pump pulse energy is 50 $\mu$J, emitted radiation wavelength is 65 nm, spontaneous life time corresponding to transition from \ket{e} to \ket{g} level is 1 ns, Auger life time is 6 fs, Auger branching ratios are taken as 0.021 for \ket{e} level and 0.0075 for \ket{g} level, pump absorption cross section is 5.2 Mb, emitted field absorption cross section is 60 Mb, gas pressure is 7 mbar, for pump a Gaussian temporal shape is assumed with 80 fs FWHM. In order to account for the spatial pulse profile of a focused beam, we modeled the intensity as a Gaussian beam with 2.0 mm Rayleigh range and 61 $\mu$m waist size. The population inversion in this system is created by Auger decay of a 4$d$ hole state, which is in turn created by photoionization due to the FEL pump \citep{1986'Kapteyn}. In terms of equations (\ref{eqs: cont LA ee}) - (\ref{eqs: cont LA G}), this scheme corresponds to the inclusion of the following rates

\begin{figure}[tb]
\includegraphics[width=1\linewidth, scale=1]{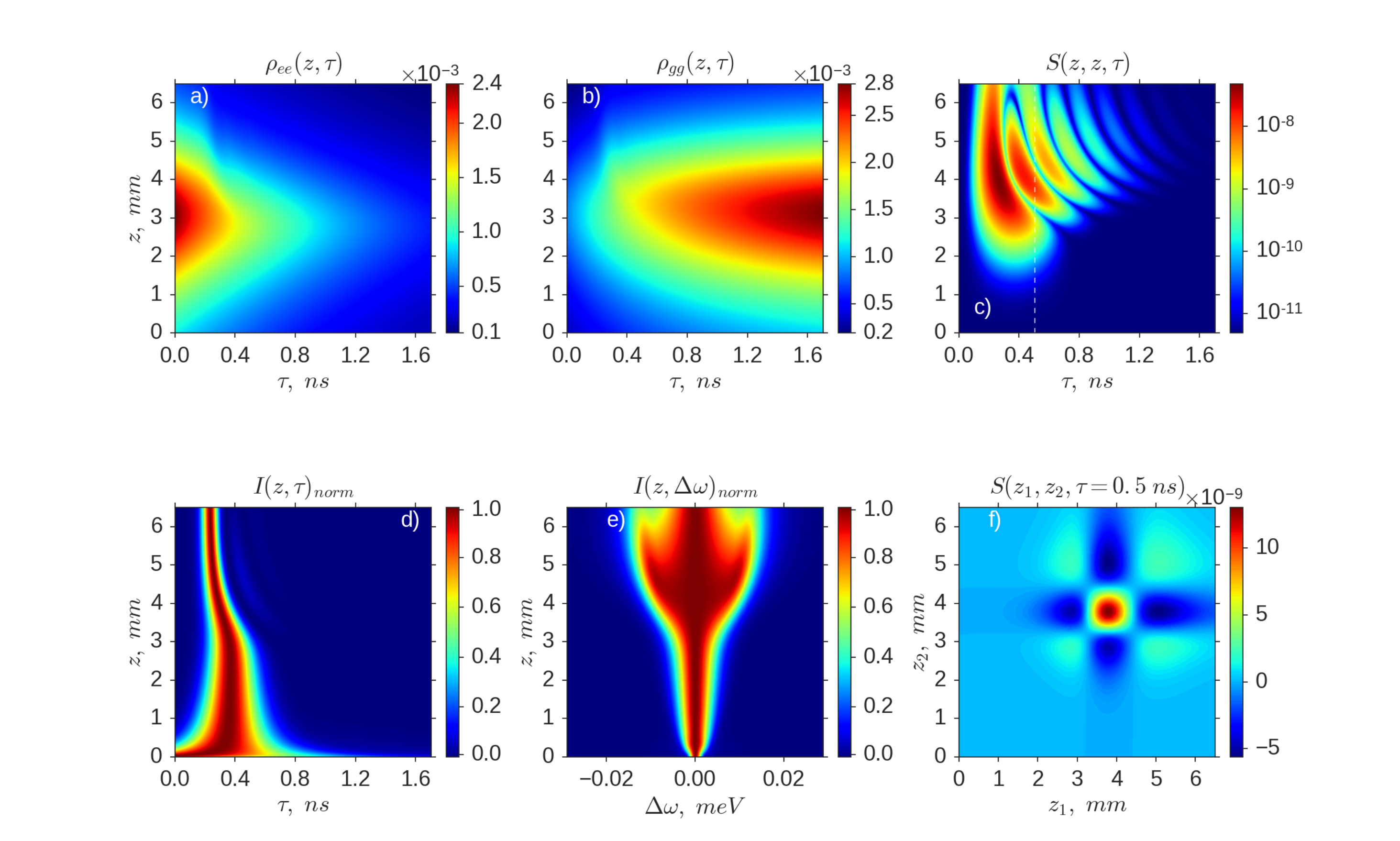}%
\caption{\label{fig: fig_CF_Xe_example} Evolution of collective spontaneous emission induced by an XFEL pump in Xe gas.  a,b - evolution of the excited and ground state populations, c -  temporal evolution of the correlation function of the atomic coherences taken at close spatial points $z_1 = z_2=z$, f - correlation function of the atomic coherences at the time moment $\tau=0.5ns$, d - profile of emitted superfluorescence intensity normalized to 1 for each $z$ value, e - spatial evolution of emitted radiation spectrum, normalized to 1 for each $z$ value.}
\end{figure}

\begin{align}
\label{eqs: rates Xe}
r_{e,g}(z,\tau)&=b_{e,g} \gamma_{A} \rho_{c}(z,\tau), \\
\gamma_{e,g}(z,\tau)&=J(z,\tau)\sigma_{i}.
\end{align}

\noindent Here, $\gamma_{A}$ is the Auger decay rate; $\rho_{c}(z,\tau)$ marks the population of the 4$d$ state, which can be obtained from rate equations for the population of atomic levels; $b_{e,g}$ are the branching ratios of the Auger decay into the levels \ket{e} and \ket{g}; $J(z,\tau)$ gives the photon flux of the FEL. It is time dependent due to the pulse shape and spatially varying due to the position-dependent beam waist size and  absorption; $\sigma_{i}$ denotes the cross section for ionization from the upper and lower level. Inclusion of the dynamics of atomic occupations  described by rate equations and the  spatial profile of the pump is obligatory for the description of realistic systems. Notably, it would be difficult to perform a similar analysis as in the Appendix \ref{appendix: Estimation of noise-term correlation factor for Maxwell-Bloch equations} and thereby obtain plausible effective noise terms for Maxwell-Bloch equations under these conditions.

The temporal field intensity  profile, Fig. \ref{fig: fig_CF_Xe_example} d,  shows transformation from spontaneous emission with maxima at $\tau =0$ to peak-shaped temporal profiles that are characteristic for superfluorescence. The peak appears earlier and becomes narrower and more intense with increasing the propagation length. 

The field correlation function enables to obtain radiation spectra by means of Fourier transform \cite{1997'Scully}. For parameters used for modeling Fig. \ref{fig: fig_CF_Xe_example}, one can see that the occupation of the \ket{e} and \ket{g} levels is of the order of $10^{-3}$, nevertheless after about 3 mm of propagation the ringing (or Rabi-like behavior) takes place, see Fig. \ref{fig: fig_CF_Xe_example} a,b. It is hardly seen in occupations, however the correlation of atomic coherences, Fig. \ref{fig: fig_CF_Xe_example} c, clearly show the oscillatory behavior in time. This behavior is present in field correlation function as well, it results in spectral broadening clearly seen in Fig. \ref{fig: fig_CF_Xe_example} e. This spectral broadening can be used as a marker for the oscillatory superfluorescence regime, and can be particularly helpful when the direct measurement of  the temporal pulse profile is not possible that is often the case for x-ray / XUV domain. At time moments corresponding to ringing, the correlation function of atomic coherences shows oscillatory behavior in space as well, see Fig. \ref{fig: fig_CF_Xe_example} f. The number of oppositely correlated regions in space (regions with opposite sign of correlation function of atomic coherences) increases with time, together with decreasing of correlation function of atomic coherences it results in damping of intensity ringing.

\section{\label{sec: Conclusions} Conclusions}
We have proposed a theoretical approach to collective spontaneous emission  that is based on integro-differential equations for the correlation function of atomic coherences and field correlation function. It enables us to obtain the spatio-temporal profiles of level occupations and emitted intensity as well as the spatial profile of the emitted radiation spectrum. Our derivations start from a basic atom-field Hamiltonian; the main approximations are the 1D modeling of the field propagation and the swept pumping configuration. A number of incoherent processes that can take place in realistic systems, such as collisional quenching of coherences, time and space dependent pumping/depleting of the levels of interest or non-radiative decay are taken into account systematically by means of the Lindblad formalism. Our approach allows for the treatment of spontaneous emission and coherent phenomena ---typically described by Maxwell-Bloch equations ---on the same footing. This gives advantage over the widely used approach of modeling collective spontaneous emission in the Maxwell-Bloch equations by means of noise terms. In particular, we show the latter one to be unable to  describe the temporal profile of spontaneous emission correctly. As an application of our method, we presented calculations corresponding to a recent experiment \cite{2018'Mercadier}. 

\begin{acknowledgments}
We gratefully acknowledge a critical reading of the manuscript by Dietrich Krebs. 
\end{acknowledgments}

\appendix
\section{\label{appendix: derivation H-L} Derivation of Heisenberg -Langevin equations}

From the Hamiltonian  (\ref{eqs: H}) we obtain the following Heisenberg equations:

\begin{align}
\label{eqs: H aSmSz:a}
\frac{d\hat{a}_{\vec{k},s}}{dt} & =-i\omega_{\vec{k}}\hat{a}_{\vec{k},s}-i g_{\vec{k},s}^*\sum_{a}e^{-i\vec{k}\vec{r_{a}}}\hat{\sigma}_{-}^{(a)},\\ 
\label{eqs: H aSmSz:Sm}
\frac{d\hat{\sigma}_{-}^{(a)}}{dt} & =-i\Omega\hat{\sigma}_{-}^{(a)}+2i\sum_{\vec{k},s}g_{\vec{k},s}e^{i\vec{k}\vec{r_{a}}}\hat{\sigma}_{z}^{(a)}\hat{a}_{\vec{k},s},\\ 
\label{eqs: H aSmSz:Sz}
\frac{d\hat{\sigma}_{z}^{(a)}}{dt} & =i\sum_{\vec{k},s}(g_{\vec{k},s}^{*}e^{-i\vec{k}\vec{r_{a}}}\hat{a}_{\vec{k},s}^{\dagger}\hat{\sigma}_{-}^{(a)}-g_{\vec{k},s}e^{i\vec{k}\vec{r_{a}}}\hat{\sigma                                                                                                                                                                                                                                                                                                                                                                                                                                                                                                                                                                                                                                                                                                                                                                                                                                                                                                                                                                                                                                                                                                                                                                                                                                                                                                                                                                                                                                                                                                                                                                                                                                                                                                                                                                                                                                                                                                                                                                                                                                                                                                                                                                                                                                                                                                                                                                                                                                                                                                                                                                                                                                                                                                                                                                                                                                                                                                                                        }_{+}^{(a)}\hat{a}_{\vec{k},s}).
\end{align}

\noindent Here, we use the normal ordering of operators: on the left side come operators $\hat{a}_{\vec{k},s}^{\dagger},\hat{\sigma}_{+}^{(a)}$ followed by $\hat{\sigma}_{z}^{(a)}$ and on the right side $\hat{a}_{\vec{k},s},\hat{\sigma}_{-}^{(a)}$, see discussions in \cite{1974'Agarwal}, \cite{1987'Allen}.

We formally integrate the equation for field operators (\ref{eqs: H aSmSz:a})

\begin{align}
\hat{a}_{\vec{k},s}(t) & =\hat{a}_{\vec{k},s}(0)e^{-i\omega_{\vec{k}}t}-ig_{\vec{k},s}^{*}\sum_{a}e^{-i\vec{k}\vec{r_{a}}}\int_{0}^{t}dt'e^{-i\omega_{\vec{k}}(t-t')}\hat{\sigma}_{-}^{(a)}(t'),
\end{align}

\noindent and substitute it into (\ref{eqs: H aSmSz:Sm}):

\begin{align}
\label{eqs: Sm expl}
\frac{d\hat{\sigma}_{-}^{(a)}}{dt}&=-i\Omega\hat{\sigma}_{-}^{(a)}+2\sum_{b,\vec{k},s}|g_{\vec{k},s}|^2\int_{0}^{t}dt'e^{i(\vec{k}\vec{(r_{a}}-\vec{r_{b}})-\omega(t-t'))}\hat{\sigma}_{z}^{(a)}(t)\hat{\sigma}_{-}^{(b)}(t') \\ \nonumber
+&2 i \sum_{\vec{k},s}g_{\vec{k},s}\hat{\sigma}_{z}^{(a)}(t)\hat{a}_{\vec{k},s}(0)e^{i(\vec{k}\vec{r_{a}}-\omega t)}.
\end{align}

\noindent The last term is determined by the vacuum field at the initial time and can be interpreted as a stochastic Langevin term 

\begin{align}
\label{eqs: def F}
\hat{F}_{-}^{(a)}(t) & =2i\sum_{\vec{k},s}g_{\vec{k},s}\hat{\sigma}_{z}^{(a)}(t)\hat{a}_{\vec{k},s}(0)e^{i(\vec{k}\vec{r_{a}}-\omega t)},\\ \nonumber
\hat{F}_{+}^{(a)}(t) & =-2i\sum_{\vec{k},s}g_{\vec{k},s}^{*}\hat{a}_{\vec{k},s}^{\dagger}(0)\hat{\sigma}_{z}^{(a)}(t)e^{-i(\vec{k}\vec{r_{a}}-\omega t)}.
\end{align}

\noindent Here $\hat{F}_{+}^{(a)}(t)$ corresponds to the analogous term in the equation for $\hat{\sigma}_{+}^{(a)}$. The statistical properties of these terms can be obtained directly from their definition (\ref{eqs: def F}) taking into account that the field is initially in the vacuum state:

\begin{align}
\label{eqs: FpFm}
\langle\hat{F}_{-}^{(a)}(t)\rangle & =\langle\hat{F}_{+}^{(a)}(t)\rangle=0, \\ \nonumber
\langle\hat{F}_{-}^{(a)}(t)\hat{F}_{-}^{(b)}(t')\rangle & =\langle\hat{F}_{+}^{(a)}(t)\hat{F}_{+}^{(b)}(t')\rangle=\langle\hat{F}_{+}^{(a)}(t)\hat{F}_{-}^{(b)}(t')\rangle=0,\\ \nonumber
\langle\hat{F}_{-}^{(a)}(t)\hat{F}_{+}^{(b)}(t')\rangle&=4\langle\hat{\sigma}_{z}^{(a)}(t)\hat{\sigma}_{z}^{(b)}(t')\rangle\sum_{\vec{k},s}|g_{\vec{k},s}|^{2}e^{i(\vec{k}(\vec{r_{a}}-\vec{r_{b}})-\omega(t-t'))}.
\end{align}

The sum over atoms in the second term of (\ref{eqs: Sm expl}) can be split into a part $b=a$ that describes the action of the field emitted by the atom $a$ on itself, and a part $b\neq a$ that describes the interaction of a given atom $a$ with other atoms in the system. Considering the first term, one can expect the action of the field emitted by atom $a$ onto itself results in spontaneous decay:

\begin{align}
\label{eqs: Sm sp}
\frac{d\hat{\tilde{\sigma}}_{-}^{(a)}}{dt}\Bigr|_{\text{sp.}} & =2\sum_{\vec{k},s}|g_{\vec{k},s}|^2\int_{0}^{t}dt'e^{-i(\omega-\Omega)(t-t')}\hat{\sigma}_{z}^{(a)}(t)\hat{\tilde{\sigma}}_{-}^{(a)}(t')\\ \nonumber
&=\frac{e^{2}|\vec{p}|^{2}}{3\pi^{2}m^{2}c^{3}\hbar\epsilon_{0}}\hat{\sigma}_{z}^{(a)}(t)\int_{0}^{t}dt'\int_{0}^{\infty}d\omega\omega\hat{\tilde{\sigma}}_{-}^{(a)}(t')e^{-i(\omega-\Omega)(t-t')} \simeq -\frac{\Gamma_{sp}}{2}\hat{\tilde{\sigma}}_{-}^{(a)}(t),\\ 
\label{eqs: Gsp def}
&\Gamma_{sp}=\frac{e^2\Omega |\vec p|^2}{3\pi\epsilon_0\hbar m^2 c^3}.
\end{align}

\noindent Here, we made use of slowly varying function $\tilde{\sigma}$: $\hat{\sigma}_{-}^{(a)}(t)=\hat{\tilde{\sigma}}_{-}^{(a)}(t)e^{-i \Omega t}$; in the last step we applied the approximation typically done in Weisskopf-Wigner approach, that is, we treat the spontaneous
emission of a single atom as a Markovian process; $\Gamma_{sp}$ is the corresponding spontaneous emission rate, where we neglected the Lamb shift. Note that the accepted rule of ordering the operators results in the description of spontaneous emission as an effect of radiation reaction alone. Conversely, other ordering conventions would result in contributions from vacuum fluctuations of the field, see \cite{1987'Allen}, Ch.7  for further discussion and references.

Next, we consider the induced part of the second term in (\ref{eqs: Sm expl}). This is responsible for the interaction of a given atom $a$ with other atoms in the system. It can be rewritten in terms of the vector potential of the electromagnetic field at the position of atom $a$ due to all other atoms in the system: 

\begin{align}
\label{eqs: Sm ind}
\frac{d\hat{\sigma}_{-}^{(a)}}{dt}\Bigr|_{\text{ind.}} & =\frac{2 i e}{m \hbar c}\hat{\sigma}_{z}^{(a)}(t)\vec{p}\cdot\vec{\hat{A}}_{+}^{(a)}(t), \\ \nonumber
\vec{\hat{A}}_{+}^{(a)}(t)&=-\frac{iec}{16\pi^{3}m\epsilon_{0}}\int d^3 \vec{k}\sum_{s}\frac{1}{\omega_{\vec{k}}}\vec{e}_{\vec{k},s}(\vec{e}_{\vec{k},s}^{*}\cdot\vec{p}^{*})\int_{0}^{t}dt'\sum_{b \neq a}e^{i(\vec{k}\vec{(r_{a}}-\vec{r_{b}})-\omega(t-t'))}\hat{\sigma}_{-}^{(b)}(t').
\end{align}

\noindent Combining (\ref{eqs: Sm sp}) and (\ref{eqs: Sm ind}) we obtain (\ref{eqs: H-L general Sm}). Similar steps lead to (\ref{eqs: H-L general Sz}), the corresponding Langevin term having zero mean value:

\begin{align}
\label{eqs: Fz}
\langle\hat{F}_{z}^{(a)}(t)\rangle = 0.
\end{align}

\section{\label{appendix: Equation of motion for product of operators} Equation of motion for the product of operators}

The equation of motion for the product of two operators can be obtained from the Heisenberg - Langevin equations of motion for each of the operators with help of generalized Einstein relations \cite{1997'Scully}. Namely, if two operators satisfy Langevin equations with $\delta$-correlated noise:

\begin{align}
\frac{d\hat{O}_{1}}{dt} & =\hat{V}_{1}(t)+\hat{F}_{1}(t), \\ \nonumber
\frac{d\hat{O}_{2}}{dt} & =\hat{V}_{2}(t)+\hat{F}_{2}(t), \\ \nonumber
\langle\hat{F}_{1}(t)\rangle & =\langle\hat{F}_{2}(t)\rangle=0,\langle\hat{F}_{1}(t)\hat{F}_{2}(t')\rangle=D_{12}(t)\delta(t-t'),
\end{align}

\noindent then observable $\langle\hat{O}_{1}(t)\hat{O}_{2}(t)\rangle$ will satisfy (see \cite{1997'Scully}, Ch. 9):

\begin{align}
\label{eqs: def Einstein relation}
\frac{d\langle\hat{O}_{1}(t)\hat{O}_{2}(t)\rangle}{dt} & =\langle\hat{V}_{1}(t)\hat{O}_{2}(t)\rangle+\langle\hat{O}_{1}(t)\hat{V}_{2}(t)\rangle+D_{12}(t).
\end{align}

Applying this expression to (\ref{eqs: HL Sm}) and its Hermitian conjugate, while taking into account (\ref{eqs: FpFm}), we obtain (\ref{eqs: mean SpSm v0}).

\section{\label{appendix: Derivation of field correlation function propagation} Derivation of the propagation equation for the field correlation function}

In order to obtain a spatial propagation equation for $G^{(a)}(\tau_{1},\tau_{2})$ we consider the difference between the field correlation function at position $z_a$ of atom $a$ and neighboring atom $a+1$ at $z_{a+1}$. From (\ref{eqs: mean G za}) we directly obtain

\begin{align}
\label{eqs: mean dG v0}
G^{(a+1)}(\tau_{1},\tau_{2})-&G^{(a)}(\tau_{1},\tau_{2}) =\\ \nonumber
&\frac{3\Delta o}{32\pi\lambda^{2}}\Gamma_{sp}\left(\sum_{b<a}\langle\hat{\sigma}_{+}^{(b)}(\tau_{1})\hat{\sigma}_{-}^{(a)}(\tau_{2})\rangle+\sum_{b<a}\langle\hat{\sigma}_{+}^{(a)}(\tau_{1})\hat{\sigma}_{-}^{(b)}(\tau_{2})\rangle+\langle\hat{\sigma}_{+}^{(a)}(\tau_{1})\hat{\sigma}_{-}^{(a)}(\tau_{2})\rangle\right).
\end{align}

\noindent With the help of (\ref{eqs: A from S}), we can express quantities like $\sum_{b<a}\hat{\sigma}_{+}^{(b)}(\tau_{1}),\,\sum_{b<a}\hat{\sigma}_{-}^{(b)}(\tau_{2})$ in terms of the field and thus find:

\begin{align}
\label{eqs: mean dG v1}
G^{(a+1)}(\tau_{1},\tau_{2})-&G^{(a)}(\tau_{1},\tau_{2}) =\\ \nonumber
&\frac{ie}{2\hbar cm\lambda^{2}}\left(p\langle\hat{\sigma}_{+}^{(a)}(\tau_{1})\hat{A}_{+}^{(a)}(\tau_{2})\rangle-p^{*}\langle\hat{A}_{-}^{(a)}(\tau_{1})\hat{\sigma}_{-}^{(a)}(\tau_{2})\rangle\right)+\frac{3\Delta o}{32\pi\lambda^{2}}\Gamma_{sp}\langle\hat{\sigma}_{+}^{(a)}(\tau_{1})\hat{\sigma}_{-}^{(a)}(\tau_{2})\rangle.
\end{align}

\noindent Next, we shall express atomic coherences in terms of the field. This can be achieved by formal integration of (\ref{eqs: HL Sm}):

\begin{align}
\label{eqs: int Sm}
\hat{\sigma}_{-}^{(a)}(\tau)=\int\limits _{0}^{\tau}d\tau'e^{-(i\Omega+\frac{\Gamma_{sp}}{2})(\tau-\tau')}\left(\frac{2 i e p}{\hbar m c}\hat{\sigma}_{z}^{(a)}(\tau')\hat{A}_{+}^{(a)}(\tau')+\hat{F}_{-}^{(a)}(\tau')\right)+e^{-(i\Omega+\frac{\Gamma_{sp}}{2})\tau}\hat{\sigma}_{-}^{(a)}(0).
\end{align}

\noindent Substituting (\ref{eqs: int Sm}) to (\ref{eqs: mean dG v1}) we obtain

\begin{align}
\label{eqs: mean dG v2}
G^{(a+1)}(\tau_{1},\tau_{2})-G^{(a)}(\tau_{1},\tau_{2})  =
\frac{3\epsilon_{0}\Omega}{4\pi\hbar c}\Gamma_{sp}&\left(\int\limits _{0}^{\tau_{1}}d\tau_{1}'e^{-(-i\Omega+\frac{\Gamma_{sp}}{2})(\tau_{1}-\tau_{1}')}\langle\hat{A}_{-}^{(a)}(\tau_{1}')\hat{\sigma}_{z}^{(a)}(\tau_{1}')\hat{A}_{+}^{(a)}(\tau_{2})\rangle\right.\\ \nonumber
&\left.+\int\limits _{0}^{\tau_{2}}d\tau_{2}'e^{-(i\Omega+\frac{\Gamma_{sp}}{2})(\tau_{2}-\tau_{2}')}\langle\hat{A}_{-}^{(a)}(\tau_{1})\hat{\sigma}_{z}^{(a)}(\tau_{2}')\hat{A}_{+}^{(a)}(\tau_{2}')\right)\\ \nonumber
+\frac{3\Delta o}{32\pi\lambda^{2}}&\Gamma_{sp}\langle\hat{\sigma}_{+}^{(a)}(\tau_{1})\hat{\sigma}_{-}^{(a)}(\tau_{2})\rangle.
\end{align}

\noindent Here, we took into account (\ref{eqs: FpFm}) as well as $\langle\hat{\sigma}_{+}^{(a)}(0)\hat{F}_{-}^{(b)}(t')\rangle=0$, $\langle\hat{F}_{+}^{(a)}(t)\hat{\sigma}_{-}^{(b)}(0)\rangle=0$, $\langle\hat{\sigma}_{+}^{(a)}(0)\hat{\sigma}_{-}^{(b)}(0)\rangle=0$ for $a\neq b$. Making approximation (\ref{eqs: approx triple factr}), which enables us to factor out $\langle\hat{\sigma}_{z}^{(a)}(\tau)\rangle$, we arrive at

\begin{align}
\label{eqs: mean dGdz all}
&G^{(a+1)}(\tau_{1},\tau_{2})-G^{(a)}(\tau_{1},\tau_{2})  = \delta G^{(a)}(\tau_{1},\tau_{2})_{\text{stim.}}+\delta G^{(a)}(\tau_{1},\tau_{2})_{\text{sp.}},\\
\label{eqs: mean dGdz stim} 
&\delta G^{(a)}(\tau_{1},\tau_{2})_{\text{stim.}}=\\ \nonumber
&\frac{3\Delta o}{8\pi}\Gamma_{sp} \left(\int\limits _{0}^{\tau_{1}}d\tau_{1}'e^{-(-i\Omega+\frac{\Gamma_{sp}}{2})(\tau_{1}-\tau_{1}')}\langle\hat{\sigma}_{z}^{(a)}(\tau_{1}')\rangle G^{(a)}(\tau_{1}',\tau_{2})+\int\limits _{0}^{\tau_{2}}d\tau_{2}'e^{-(i\Omega+\frac{\Gamma_{sp}}{2})(\tau_{2}-\tau_{2}')}\langle\hat{\sigma}_{z}^{(a)}(\tau_{2}')\rangle G^{(a)}(\tau_{1},\tau_{2}')\right),\\ 
\label{eqs: mean dGdz sp}
&\delta G^{(a)}(\tau_{1},\tau_{2})_{\text{sp.}}=\frac{3\Delta o}{32\pi\lambda^{2}}\Gamma_{sp} \langle\hat{\sigma}_{+}^{(a)}(\tau_{1})\hat{\sigma}_{-}^{(a)}(\tau_{2})\rangle.
\end{align}

\noindent Here, (\ref{eqs: mean dGdz stim}) describes the contribution to the field due to stimulated emission of atom $a$. This results in amplification for the case when the atom is in an inverted state $\langle\hat{\sigma}_{z}^{(a)}\rangle > 0$ and in absorption otherwise; (\ref{eqs: mean dGdz sp}) describes the contribution to the field due to spontaneous emission of atom $a$. Evaluation of the spontaneous emission requires the knowledge of the one-atom two-time correlation function for atomic coherences. With the help of (\ref{eqs: int Sm}), we can express it as

\begin{align}
\label{eqs: mean Spt1Smt2 all}
\langle\hat{\sigma}_{+}^{(a)}(\tau_{1})\hat{\sigma}_{-}^{(a)}(\tau_{2})\rangle&=\langle\hat{\sigma}_{+}^{(a)}(\tau_{1})\hat{\sigma}_{-}^{(a)}(\tau_{2})\rangle_{\text{free}}+\langle\hat{\sigma}_{+}^{(a)}(\tau_{1})\hat{\sigma}_{-}^{(a)}(\tau_{2})\rangle_{\text{stim.}},\\ 
\label{eqs: mean Spt1Smt2 free}
\langle\hat{\sigma}_{+}^{(a)}(\tau_{1})\hat{\sigma}_{-}^{(a)}(\tau_{2})\rangle_{\text{free}}&=\rho_{ee}^{(a)}(0)e^{i\Omega(\tau_{1}-\tau_{2})}e^{-\frac{\Gamma_{sp}}{2}(\tau_{1}+\tau_{2})}, \\
\label{eqs: mean Spt1Smt2 stim}
\langle\hat{\sigma}_{+}^{(a)}(\tau_{1})\hat{\sigma}_{-}^{(a)}(\tau_{2})\rangle_{\text{stim.}}&=\\ \nonumber
\frac{3\Delta o\lambda^{2}}{2\pi}\Gamma_{sp}\int\limits _{0}^{\tau_{1}}d\tau_{1}'\int\limits _{0}^{\tau_{2}}d\tau_{2}' & e^{i\Omega(\tau_{1}-\tau_{1}'-[\tau_{2}-\tau_{2}'])}e^{-\frac{\Gamma_{sp}}{2}(\tau_{1}-\tau_{1}'+\tau_{2}-\tau_{2}')}\langle\hat{\sigma}_{z}^{(a)}(\tau_{1}')\hat{\sigma}_{z}^{(a)}(\tau_{2}')\rangle G^{(a)}(\tau_{1}',\tau_{2}').
\end{align}

\noindent Here, (\ref{eqs: mean Spt1Smt2 free}) gives the value of $\langle\hat{\sigma}_{+}^{(a)}(\tau_{1})\hat{\sigma}_{-}^{(a)}(\tau_{2})\rangle$ in the case of a free atom unaffected by any external field, while (\ref{eqs: mean Spt1Smt2 stim}) provides a correction due to the exposure to a field with a given correlation function. Exact evaluation of (\ref{eqs: mean Spt1Smt2 stim}) would require the knowledge of the single-atom two-time quantity $\langle\hat{\sigma}_{z}^{(a)}(\tau_{1}')\hat{\sigma}_{z}^{(a)}(\tau_{2}')\rangle$, which would in turn need a separate set of equations. However, the contribution of this term only becomes important, once the field reaches significant value, at which point the stimulated contribution (\ref{eqs: mean dGdz stim}) already dominate over the spontaneous term of (\ref{eqs: mean dGdz sp}). Hence, there is no need to calculate the correction (\ref{eqs: mean Spt1Smt2 stim}) to an anyway negligible term (\ref{eqs: mean dGdz sp}). 

The same conclusion can be drawn from the following quantitative argument. Let us make an order-of-magnitude estimate for the terms (\ref{eqs: mean dGdz sp}),(\ref{eqs: mean dGdz stim}),(\ref{eqs: mean Spt1Smt2 free}), (\ref{eqs: mean Spt1Smt2 stim}):

\begin{align}
\label{eqs: mean dGdz stim est} 
\delta G_{\text{stim.}} &\sim \Delta o \Gamma_{sp} \tau G \rho_{inv}, \\
\label{eqs: mean dGdz sp est}
\delta G_{\text{sp.}} &\sim \frac{\Delta o}{\lambda^{2}}\Gamma_{sp}\rho_{ee}, \\
\label{eqs: mean Spt1Smt2 stim est}
\langle\hat{\sigma}_{+}\hat{\sigma}_{-}\rangle_{\text{stim.}} &\sim \Delta o\lambda^{2} \Gamma_{sp} \tau^2 G \rho_{inv}^2, \\
\label{eqs: mean Spt1Smt2 free est}
\langle\hat{\sigma}_{+}\hat{\sigma}_{-}\rangle_{\text{sp.}} &\sim\rho_{ee}.
\end{align}

\noindent Here, $G$, $\tau$,  $\rho_{inv}$   are typical values of the field correlation function, evolution time and population inversion, respectively. From (\ref{eqs: mean dGdz stim est}) - (\ref{eqs: mean Spt1Smt2 free est}) we obtain the relation:

\begin{align}
\label{eqs: est stim to sp ratio}
\frac{\langle\hat{\sigma}_{+}\hat{\sigma}_{-}\rangle_{\text{stim.}}}{\langle\hat{\sigma}_{+}\hat{\sigma}_{-}\rangle_{\text{sp.}}} \sim \frac{\delta G_{\text{stim.}}}{\delta G_{\text{sp.}}} \Delta o \rho_{inv} \Gamma_{sp} \tau.
\end{align}

\noindent Accurate calculation of (\ref{eqs: mean Spt1Smt2 stim}) is important only while (\ref{eqs: mean dGdz stim}) is smaller than (\ref{eqs: mean dGdz sp}), hence we can set 

\begin{align}
\label{eqs: est stim to sp less 1}
\frac{\delta G_{\text{stim.}}}{\delta G_{\text{sp.}}} \lesssim 1
\end{align}

\noindent in (\ref{eqs: est stim to sp less 1}). The typical time to build-up the field that would dominate over the spontaneous emission can be estimated as the superradiance time $\tau_{SR} \sim \frac{1}{\Gamma_{sp} \rho_{inv} n L \Delta o}$, here $n L$ is the number of atoms in the medium. Hence, while we assume that (\ref{eqs: est stim to sp less 1}) holds, we can take 

\begin{align}
\label{eqs: est tau}
\tau \lesssim \frac{1}{\Gamma_{sp} \rho_{inv} n L \Delta o}.
\end{align} 

\noindent In the frame of (\ref{eqs: est stim to sp less 1}) and (\ref{eqs: est tau}) we straightforwardly find with (\ref{eqs: est stim to sp ratio})

\begin{align}
\label{eqs: est result stim to sp ratio}
\frac{\langle\hat{\sigma}_{+}\hat{\sigma}_{-}\rangle_{\text{stim.}}}{\langle\hat{\sigma}_{+}\hat{\sigma}_{-}\rangle_{\text{sp.}}} \lesssim \frac{1}{n L} \ll 1.
\end{align}

\noindent This shows that in the region, where (\ref{eqs: mean dGdz sp}) is important, we can neglect the contribution of (\ref{eqs: mean Spt1Smt2 stim}) with respect to (\ref{eqs: mean Spt1Smt2 free}). This conclusion was further supported by numerical calculation of the ratio between (\ref{eqs: mean Spt1Smt2 free}) and the estimate of (\ref{eqs: mean Spt1Smt2 stim}). Another way of checking it is to calculate the intensity directly from atomic variables (\ref{eqs: I za}) and alternatively from the field correlation function $I^{(a)}(\tau)=G^{(a)}(\tau,\tau)$. In the first case, no approximation with regard to (\ref{eqs: mean Spt1Smt2 stim}) is made and its results agree with the approximate method. Hence, we will make the approximation of dropping off the contribution (\ref{eqs: mean Spt1Smt2 stim}) and arrive at (\ref{eqs: mean dGdz}).

\section{\label{appendix: Modification of equations for atomic and field observables due to incoherent processes}Modification of equations for atomic and field observables due to incoherent processes}

Here, we consider how the expressions that were used to derive (\ref{eqs: cont rinv}) - (\ref{eqs: cont G}) change in the presence of incoherent processes described by Lindblad superoperator (\ref{eqs: L tot}). 


The interaction with a Markovian reservoir can be described within the frame of Heisenberg-Langevin equations by the addition of a regular and a stochastic operator \cite{1997'Scully}, Ch.9. Hence, equations of the form (\ref{eqs: HL Sm}), (\ref{eqs: HL Sz}) are transformed into

\begin{align}
\label{eqs: def Langevin}
\frac{d\hat{O}_{\mu}^{(a)}(\tau)}{d\tau}&=\left(\frac{d\hat{O}_{\mu}^{(a)}(\tau)}{d\tau}\right)_{\text{field}}+\hat{V}_{\mu}^{(a)}(\tau)+\hat{F}_{\mu}^{(a)}(\tau),\\ 
\label{eqs: def noise Langevin}
\langle\hat{F}_{\mu}^{(a)}(\tau)\rangle&=0, \\ \nonumber
\langle\hat{F}_{\mu}^{(a)}(\tau_{1})\hat{F}_{\nu}^{(b)}(\tau_{2})\rangle&=D_{\mu\nu}(\tau_{1})\delta_{ab}\delta(\tau_{1}-\tau_{2}).
\end{align}

\noindent Here, $\hat{O}_{\mu}$ stands for some set of operators, e.g., $\sigma_+,\sigma_-,\sigma_z$; the first term comprises the evolution due to spontaneous decay and the interaction between atoms via their emitted field; the second and the third terms are regular and noise (Langevin) contributions due to the incoherent processes; the Kronecker symbol $\delta_{ab}$ in (\ref{eqs: def noise Langevin}) reflects our assumption that noise terms are uncorrelated if they correspond to different atoms (uncorrelated reservoirs).  

If one knows the solution of master equation for one-atom density matrix, the value of any one-atom observable can be obtained directly from it, see \cite{1997'Scully}, Ch. 9. Also, from the definition of quantum-mechanical mean we have

\begin{align}
\frac{d\langle\hat{O}_{\mu}^{(a)}(\tau)\rangle}{d\tau}=\left(\frac{d\langle\hat{O}_{\mu}^{(a)}(\tau)\rangle}{d\tau}\right)_{\text{field}}+\text{Tr}\left[\hat{O}_{\mu}^{(a)}(\tau)\left(\frac{\partial\hat{\rho}^{(a)}}{\partial\tau}\right)_{\text{incoh}}\right].
\end{align}

\noindent The last term can be presented as

\begin{align}
\label{eqs: L O}
\text{Tr}\left[\hat{O}_{\mu}^{(a)}(\tau)\left(\frac{\partial\hat{\rho}^{(a)}}{\partial\tau}\right)_{\text{incoh}}\right]=\text{Tr}\left[\hat{O}_{\mu}^{(a)}(\tau)L^{(a)}\{\hat{\rho}^{(a)}\}\right]=\text{Tr}\left[\hat{V}_{\mu}^{(a)}(\tau)\hat{\rho}^{(a)}\right].
\end{align}

\noindent We can now obtain $\hat{V}_{\mu}^{(a)}(\tau)$ from the last equation using the actual form of the Lindblad superoperator (\ref{eqs: L tot}) and taking into account that we can perform cyclic permutation of operators under the trace. 

\subsection*{Equations for $\rho_{ee}$, $\rho_{gg}$}

In this part, we outline how the equation for the population inversion would modify. In contrast to the case of a pure two level system considered in Sec.\ref{sec: Derivation_I}, the identity $\rho_{ee}^{(a)}+\rho_{gg}^{(a)}=1$ is not satisfied in general case. Hence we have to obtain separate equations for $\rho_{ee}^{(a)}(\tau)=\langle\hat{\sigma}_{+}^{(a)}(\tau)\hat{\sigma}_{-}^{(a)}(\tau)\rangle$ and $\rho_{gg}^{(a)}(\tau)=\langle\hat{\sigma}_{-}^{(a)}(\tau)\hat{\sigma}_{+}^{(a)}(\tau)\rangle$. Using these operators in (\ref{eqs: L O}) we obtain

\begin{align}
\label{eqs mean VeeVgg}
\langle\hat{V}_{ee}^{(a)}(\tau)\rangle&=r_e(z_a,\tau)-\left(\gamma_e(z_a,\tau)+\gamma_n\right)\rho_{ee}^{(a)}(\tau), \\ \nonumber
\langle\hat{V}_{gg}^{(a)}(\tau)\rangle&=r_g(z_a,\tau)+\gamma_{n}\rho_{ee}^{(a)}(\tau)-\gamma_{g}(z_a,\tau)\rho_{gg}^{(a)}(\tau),
\end{align}

\noindent Here, $r_{e,g}(z_a,\tau) = \tilde{r}_{e,g}(z_a,\tau) \rho_{xx}^{(a)}(\tau)$ is the pumping rate of level \ket{e} or \ket{g}, which is proportional to the occupation of the state \ket{x}, from which the pumping takes place. Taking into account that contributions due to spontaneous decay and interaction with emitted field can be expressed as

\begin{align}
\label{eqs: ee gg via sz}
\left(\frac{d\rho_{ee}^{(a)}(\tau)}{d\tau}\right)_{\text{field}}=\left(\frac{d\langle\hat{\sigma}_{z}^{(a)}(\tau)\rangle}{d\tau}\right)_{\text{field}}, \quad \left(\frac{d\rho_{gg}^{(a)}(\tau)}{d\tau}\right)_{\text{field}}=-\left(\frac{d\langle\hat{\sigma}_{z}^{(a)}(\tau)\rangle}{d\tau}\right)_{\text{field}},
\end{align}

\noindent we obtain as new equations for the continuous variables:

\begin{align}
\label{eqs: cont L ee}
\frac{\partial\rho_{ee}(z,\tau)}{\partial\tau}&=r_e(z,\tau)-\left(\Gamma_{sp}+\gamma_e(z,\tau)+\gamma_n\right)\rho_{ee}(z,\tau)-\frac{3\text{\ensuremath{\Delta o}}}{8\pi}\Gamma_{sp}n\int\limits _{0}^{z}dz'S(z,z',\tau),\\ \nonumber
&\Gamma_{ee}(z,\tau)=\Gamma_{sp}+\gamma_e(z,\tau)+\gamma_n, \\
\label{eqs: cont L gg}
\frac{\partial\rho_{gg}(z,\tau)}{\partial\tau}&=r_g(z,\tau)+(\Gamma_{sp}+\gamma_{n})\rho_{ee}(z,\tau)-\gamma_{g}(z,\tau)\rho_{gg}(z,\tau)+\frac{3\text{\ensuremath{\Delta o}}}{8\pi}\Gamma_{sp}n\int\limits _{0}^{z}dz'S(z,z',\tau).
\end{align}

\subsection*{Equations for $S(z_1,z_2,\tau)$}

From expression (\ref{eqs: L O}) applied to operators $\hat{\sigma}_{\pm}^{(a)}$, we find

\begin{align}
\label{eqs: Vpm}
\hat{V}_{\pm}^{(a)}(\tau)=-\frac{1}{2}\left[\gamma_n+q(z_a,\tau)+\gamma_e(z_a,\tau)+\gamma_{g}(z_a,\tau)\right]\hat{\sigma}_{\pm}^{(a)}(\tau).
\end{align}

\noindent Hence, in equations involving $\hat{\sigma}_{\pm}^{(a)}$ operators, such as (\ref{eqs: HL Sm}), the account of (\ref{eqs: Vpm}) is equivalent to modifying the decay term by changing the decay rate to 

\begin{align}
\label{eqs: def Gtotal}
\Gamma(z_a,\tau)=\Gamma_{sp}+\gamma_n+q(z_a,\tau)+\gamma_e(z_a,\tau)+\gamma_{g}(z_a,\tau).
\end{align}

Expression (\ref{eqs: HL Sm}) and its hermitian conjugate were used to obtain the equation for the time-propagation of the correlation function of atomic coherences (\ref{eqs: mean SpSm v0}). Taking into account that the derivation was based on quantum Einstein relation (\ref{eqs: def Einstein relation}), one can see that the inclusion of additional operators (\ref{eqs: Vpm}) results in the following substitution of the term $-\Gamma_{sp}S(z_{1},z_{2},\tau)$ in (\ref{eqs: cont S}) by

\begin{align}
\label{eqs: subst L cont S}
-\Gamma_{sp}S(z_{1},z_{2},\tau) \rightarrow -\frac{1}{2}\left[\Gamma(z_1,\tau)+\Gamma(z_2,\tau)\right]S(z_{1},z_{2},\tau).
\end{align}

Note that the term $D_{12}$ in (\ref{eqs: def Einstein relation}) gives no contribution due to our assumption that the reservoirs  are independent for each atom.

\subsection*{Equations for the field correlation function}

Finally, we discuss, which modifications need to be introduced to the derivation done in Appendix \ref{appendix: Derivation of field correlation function propagation}. In expression (\ref{eqs: int Sm}), we have to take into account the additional  terms from (\ref{eqs: Vpm}); this results in:

\begin{align}
\label{eqs: L int Sm}
\hat{\sigma}_{-}^{(a)}(\tau)=&\int\limits _{0}^{\tau}d\tau'e^{-i\Omega(\tau-\tau')}e^{-\frac{1}{2}\int_{\tau'}^{\tau}d\tau''\Gamma(z_a,\tau'')}\left(\frac{2 i e p}{\hbar m c}\hat{\sigma}_{z}^{(a)}(\tau')\hat{A}_{+}^{(a)}(\tau')+\hat{F}_{-}^{(a)}(\tau')\right) \\ \nonumber
&+e^{-i\Omega\tau}e^{-\frac{1}{2}\int_{0}^{\tau}d\tau''\Gamma(z_a,\tau'')}\hat{\sigma}_{-}^{(a)}(0),
\end{align}

\noindent Retaining the damping term $e^{-\frac{1}{2}\int_{\tau'}^{\tau}d\tau''\Gamma(z_a,\tau'')}$ throughout further derivations results in the factor $\mathcal{D}$ appearing in (\ref{eqs: cont LA G}).

Additional modifications affect the quantity $\langle\hat{\sigma}_{+}^{(a)}(\tau_{1})\hat{\sigma}_{-}^{(a)}(\tau_{2})\rangle$, which is needed to calculate (\ref{eqs: mean dGdz sp}). Following the same reasoning as in Appendix \ref{appendix: Derivation of field correlation function propagation} we consider only contributions from $\langle\hat{\sigma}_{+}^{(a)}(\tau_{1})\hat{\sigma}_{-}^{(a)}(\tau_{2})\rangle_{\text{free}}$, i.e., evolution only due to the spontaneous decay and incoherent processes of individual atoms. To this end, we use (\ref{eqs: L int Sm}) omitting the field and write

\begin{align}
\label{eqs: L Spm v0}
\langle\hat{\sigma}_{+}^{(a)}(\tau_{1})&\hat{\sigma}_{-}^{(a)}(\tau_{2})\rangle=\\ \nonumber
&\int\limits _{0}^{\tau_{1}}d\tau_{1}'\int\limits _{0}^{\tau_{2}}d\tau_{2}'e^{i\Omega(\tau_{1}-\tau_{1}'-[\tau_{2}-\tau_{2}'])}e^{-\frac{1}{2}\left[\int_{\tau_1'}^{\tau_1}d\tau_1''\Gamma(z_a,\tau_1'')+\int_{\tau_2'}^{\tau_2}d\tau_2''\Gamma(z_a,\tau_2'')\right]}
\langle\hat{F}_{+}^{(a)}(\tau_{1}')\hat{F}_{-}^{(a)}(\tau_{2}')\rangle \\ \nonumber
&+e^{-\frac{1}{2}\left[\int_{0}^{\tau_1}d\tau_1''\Gamma(z_a,\tau_1'')+\int_{0}^{\tau_2}d\tau_2''\Gamma(z_a,\tau_2'')\right]}e^{i\Omega(\tau_{1}-\tau_{2})}\rho_{ee}^{(a)}(0).
\end{align}

\noindent In order to evaluate (\ref{eqs: L Spm v0}), we need to know the value of the noise correlation $D_{+-}(\tau)$, see (\ref{eqs: def noise Langevin}). It can be obtained from the combination of the master equation and the generalized Einstein relation (\ref{eqs: def Einstein relation}), see \cite{1997'Scully} Ch. 9.4:

\begin{align}
\label{eqs: L Dpm}
D_{+-}(\tau)&=\langle\hat{V}_{ee}^{(a)}(\tau)\rangle-\langle\hat{V}_{+}^{(a)}(\tau)\hat{\sigma}_{-}^{(a)}(\tau)\rangle-\langle\hat{\sigma}_{+}^{(a)}(\tau)\hat{V}_{-}^{(a)}(\tau)\rangle\\ \nonumber
&=r_e(z_a,\tau)+\left[\Gamma(z_a,\tau)-\Gamma_{ee}(z_a,\tau)\right]\rho_{ee}^{(a)}(\tau).
\end{align}

\noindent With the help of (\ref{eqs: def noise Langevin}) and using (\ref{eqs: L Dpm}), the expression (\ref{eqs: L Spm v0}) transforms to

\begin{align}
\label{eqs: L Spm v1}
&\langle\hat{\sigma}_{+}^{(a)}(\tau_{1})\hat{\sigma}_{-}^{(a)}(\tau_{2})\rangle=\\ \nonumber
&e^{i\Omega(\tau_{1}-\tau_{2})}\int\limits _{0}^{\min{\tau_1,\tau_2}}d\tau'e^{-\frac{1}{2}\left[\int_{\tau'}^{\tau_1}d\tau_1''\Gamma(z_a,\tau_1'')+\int_{\tau'}^{\tau_2}d\tau_2''\Gamma(z_a,\tau_2'')\right]}
\left[r_e(z_a,\tau')+\left[\Gamma(z_a,\tau')-\Gamma_{ee}(z_a,\tau')\right]\rho_{ee}^{(a)}(\tau')\right] \\ \nonumber
&+e^{i\Omega(\tau_{1}-\tau_{2})}e^{-\frac{1}{2}\left[\int_{0}^{\tau_1}d\tau_1''\Gamma(z_a,\tau_1'')+\int_{0}^{\tau_2}d\tau_2''\Gamma(z_a,\tau_2'')\right]}\rho_{ee}^{(a)}(0).
\end{align}

\noindent This constitutes the spontaneous emission part of (\ref{eqs: cont LA G}).

\section{\label{appendix: Derivation of Maxwell-Bloch equations from correlation fucntion equations}Derivation of Maxwell-Bloch equations from the correlation function equations}

Here, we consider how the correlation function equations (\ref{eqs: cont LA ee}) - (\ref{eqs: cont LA G}) simplify if we assume the factorization of (\ref{eqs: MB factorization S}), (\ref{eqs: MB factorization G}) and omit the spontaneous emission terms. Under these assumptions, we can substitute (\ref{eqs: MB factorization S}) into (\ref{eqs: cont LA S}) and obtain the following expression

\begin{align}
\label{eqs: cont LA MB S eq1}
\rho_{ge}(z_1,\tau) & \frac{\partial \rho_{eg}(z_2,\tau)}{\partial\tau} + \rho_{eg}(z_2,\tau)\frac{\partial \rho_{ge}(z_1,\tau)}{\partial\tau}= \\ \nonumber
&\rho_{ge}(z_1,\tau)\left(-\frac{\Gamma(z_2,\tau)}{2}\rho_{eg}(z_2,\tau)+\frac{3\Delta o}{16\pi}\Gamma_{sp}\rho_{inv}(z_{2},\tau)n\int\limits _{0}^{z_{2}}dz_{2}'\mathcal{A}(z_2,z_2')\rho_{eg}(z_{2}',\tau)\right)\\ \nonumber
&\rho_{eg}(z_2,\tau)\left(-\frac{\Gamma(z_1,\tau)}{2}\rho_{ge}(z_1,\tau)+\frac{3\Delta o}{16\pi}\Gamma_{sp}\rho_{inv}(z_{1},\tau)n\int\limits _{0}^{z_{1}}dz_{1}'\mathcal{A}(z_1,z_1')\rho_{ge}(z_{1}',\tau)\right).
\end{align}

\noindent This itself can obviously be expressed in factorized form as well. The integral in (\ref{eqs: cont LA MB S eq1}) can be represented as the solution of a differential equation and in this way, we can introduce an electric field envelope that satisfies

\begin{align}
\label{eqs: MB E app}
\frac{\partial \mathcal{E}_{+}(z,\tau)}{\partial z} &=-\frac{\kappa(z)}{2}\mathcal{E}_{+}(z,\tau) + \frac{i \Omega}{2 \epsilon_0 c}\mu n_v \rho_{eg}(z,\tau) \cdot \xi, \\ \nonumber
\xi &= \frac{\Delta o \pi R^2}{2 \lambda^2}.
\end{align}

\noindent Using this, we obtain from (\ref{eqs: cont LA MB S eq1})

\begin{align}
\label{eqs: cont MB rho ge app}
\frac{\partial \rho_{ge}(z,\tau)}{\partial\tau}=-\frac{\Gamma(z,\tau)}{2}\rho_{ge}(z,\tau)+\frac{i \mu}{\hbar}\rho_{inv}(z,\tau)\mathcal{E}_{-}(z,\tau). 
\end{align}

\noindent Here, we used the definition of the spontaneous decay rate (\ref{eqs: Gsp def}), and the dipole moment $\mu$ that is related to the momentum matrix element given in (\ref{eqs: gks and p}) as $e\,p=i m \Omega \mu$. Furthermore, $n_v$ is the three-dimensional particle density, namely $n=n_v \pi R^2$, while $\mathcal{E}_{+}$ and $\mathcal{E}_{-}$ are the positive and negative frequency parts of the field; they are complex conjugate to one another. The way how to split the factor $3\Delta o/(16\pi)\Gamma_{sp}$ into prefactors in (\ref{eqs: MB E app}) and (\ref{eqs: cont MB rho ge app}) is arbitrary at this stage. The chosen factors comply with the expressions (\ref{eqs: H-L general Sm}), (\ref{eqs: A from S}). Since these expressions are our basic expressions to obtain field correlation function equation (\ref{eqs: cont LA G}), the field propagation equation (\ref{eqs: MB E app}) should agree with (\ref{eqs: cont LA G}), (\ref{eqs: MB factorization G}). Notably, the semi-classical Maxwell-Bloch equations could also be obtained directly from (\ref{eqs: H-L general Sm}), (\ref{eqs: H-L general Sz}), (\ref{eqs: A from S}) and (\ref{eqs: Vpm}) by replacing all operators by $c$-numbers and omitting the operator noise terms. 

The field equation (\ref{eqs: MB E app}) differs from typically used Maxwell-Bloch equations \citep{1982'Gross} by a factor $\xi$. This is result of the 1D approximation (\ref{eqs: 1D approximation}), though within this approximation, it can be taken as $\xi \sim 1$. With this choice we obtain (\ref{eqs: MB ee}) - (\ref{eqs: MB E}).

\section{\label{appendix: Estimation of noise-term correlation factor for Maxwell-Bloch equations}Noise-term correlation factor for Maxwell-Bloch equations} 

In order to compare the present approach to Maxwell-Bloch equations with noise terms we derive the noise term correlation factor $F$ that is consistent with the approximations and conventions used throughout this paper. We restrict ourselves to a simplified scheme, assuming instantaneous pumping and the absence of incoherent processes except for decoherence with a fixed decoherence rate $q$. If no external field is present, then from (\ref{eqs: cont MB rho ge app}) supplemented with the noise term $s_{+}(z,\tau)$ we obtain

\begin{align}
\label{eqs: noise eq1 rho}
\rho_{ge}(z,\tau)=\int_0^{\tau}d\tau'e^{-\frac{\Gamma}{2}(\tau-\tau')}s_{+}(z,\tau'),
\end{align}

\noindent where $\Gamma=\Gamma_{sp}+q$. Substituting this expression into (\ref{eqs: MB E app}) and considering the case without absorption, we find

\begin{align}
\label{eqs: noise eq2 E}
\mathcal{E}_{+}(z,\tau)=\frac{i \Omega \mu n_v}{2 \epsilon_0 c}\int_0^{z}dz'\int_0^{\tau}d\tau'e^{-\frac{\Gamma}{2}(\tau-\tau')}s_{-}(z',\tau'),
\end{align}

\noindent with $\xi \sim 1$.

The number of spontaneously emitted photons can be obtained as 

\begin{align}
\label{eqs: noise eq3 Nsp}
\frac{d N_{\text{ph.,sp.}}(\tau)}{d\tau}=\frac{2 \epsilon_0 c}{\hbar \omega}\langle \mathcal{E}_{-}(\tau)\mathcal{E}_{+}(\tau) \rangle \pi R^2,
\end{align}

\noindent where $\langle..\rangle$ indicates the average over realizations of the stochastic variables $s_{\pm}(z,\tau)$. Substituting (\ref{eqs: noise eq2 E})  into (\ref{eqs: noise eq3 Nsp}), taking into account (\ref{eqs: s eq1}), performing integrals with the $\delta$-functions and using expression (\ref{eqs: Gsp def}), we arrive at
\begin{align}
\label{eqs: noise eq4 Nsp int}
\frac{d N_{\text{ph.,sp.}}(\tau)}{d\tau}=\frac{3}{8 \pi}\Gamma_{sp} n z n_v F \lambda^2  \int_0^{\tau}d\tau'e^{-\Gamma (\tau-\tau')}\rho_{ee}(\tau').
\end{align}

We can now compare the obtained expression with the exact result (\ref{eqs: Nsp exact}). If we assume that the decoherence time is much shorter than the spontaneous emission time, we can take  $\rho_{ee}(\tau)$ out of the integral and obtain

\begin{align}
\label{eqs: noise eq5 F}
F=\frac{2 \Gamma}{n}\cdot \xi.
\end{align}

\noindent Here again, we can take $\xi \sim 1$. However, if $\Gamma=\Gamma_{sp}$ then within the choice (\ref{eqs: noise eq5 F}) for the temporal dependence of spontaneously emitted quanta we find

\begin{align}
\label{eqs: noise eq6 Nsp for G is Gsp}
\frac{d N_{\text{ph.,sp.}}(\tau)}{d\tau}=\frac{3}{8 \pi}\Gamma_{sp} n z \Delta o \rho_{ee}(0) \Gamma_{sp}\tau e^{-\Gamma_{sp} \tau},
\end{align}

\noindent under the same approximation $\xi \sim 1$. Clearly, (\ref{eqs: noise eq6 Nsp for G is Gsp}) differs from the correct result by a factor $\Gamma_{sp}\tau$.


%

\end{document}